\documentclass[aps,prd,superscriptaddress,showpacs,preprintnumbers]{revtex4}
\usepackage{graphicx}

\begin{document}


\parskip=0.3cm


\title{Low missing mass, single- and double diffraction dissociation at the LHC}

\author{L\'aszl\'o Jenkovszky}
\affiliation{Bogolyubov Institute for Theoretical Physics (BITP),
Ukrainian National Academy of Sciences \\14-b, Metrolohichna str.,
Kiev, 03680, Ukraine; \\Wigner Research Centre for Physics, Hungarian Academy of Sciences\\
1525 Budapest, POB 49, Hungary}

\author{Oleg Kuprash}
\affiliation{Deutsches Elektronen-Synchrotron, Notkestraße 85, 22607 Hamburg, Germany}

\author{Risto Orava}
\affiliation{Helsinki
Institute of Physics, Division of Elementary Particle Physics, P.O. Box 64 (Gustaf H\"allstr\"ominkatu 2a),
FI-00014 University of Helsinki, Finland; \\CERN, CH-1211 Geneva 23, Switzerland}

\author{Andrii Salii}
\affiliation{Bogolyubov Institute for Theoretical Physics (BITP),
Ukrainian National Academy of Sciences \\14-b, Metrolohichna str.,
Kiev, 03680, Ukraine}

\begin{abstract}
Low missing mass, single- and double diffraction dissociation is calculated for the LHC energies from a dual-Regge model, dominated by a Pomeron Regge pole exchange. The model reproduces the rich resonance structure in the low missing mass $M_X$ region. The diffractively excited states lie on the nucleon trajectory $N^*$, appended by the isolated Roper resonance. Detailed predictions for the squared momentum transfer and missing mass dependence of the differential and integrated single- and double diffraction dissociation in the kinematical range of present and future LHC measurements are given.

  The present work is a continuation and extension (e.g. with double diffraction) of a previous work \cite{PR} using the dual Regge approach.

\end{abstract}

\pacs{11.55.-m, 11.55.Jy, 12.40.Nn}

\maketitle


\section{Introduction}
Measurements of single (\textbf{SD}), double(\textbf{DD}) and central(\textbf{CD}) diffraction dissociation is among the priorities of the LHC research program.

In the past, intensive studies of high-energy diffraction dissociation were performed at the Fermilab,
on fixed deuteron target, and at the ISR, see \cite{Goulianos} for an overview and relevant references. Fig.~\ref{fig:2SD.Fermilab}
shows representative curves of low-mass SD as measured at the Fermilab. One can see the rich resonance structure there,
typical for low missing masses, often ignored by extrapolating whole region by a simple $1/M^2$ dependence.
 When extrapolating (in energy), one should however bear in mind that, in the ISR region, secondary Reggeon contributions are still important (their relative contribution depends on momenta transfer considered), amounting to nearly $50\%$ in the forward direction. At the LHC, however, their contribution in the nearly forward direction in negligible, i.e. less than the relevant error bars in the measured total cross section \cite{JLL}.

\begin{figure}[!ht]
 \centering
  \includegraphics[width=0.8\textwidth ,bb= 0 0 730 180]{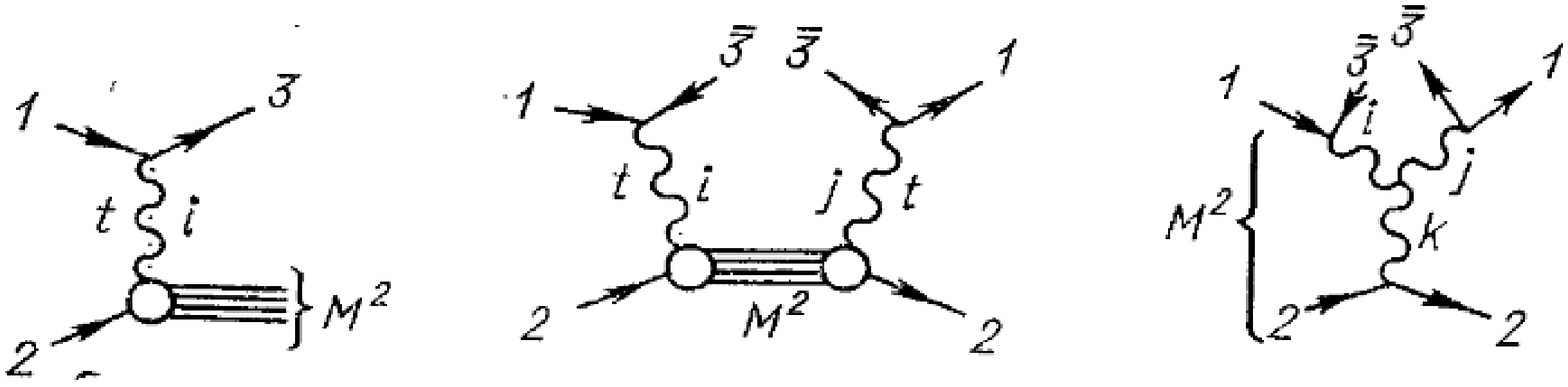}
  \caption{From SD to the triple Regge limit.}
  \label{fig:TripleReggeLimit}
 \end{figure}

 In most of the papers on the subject SD is calculated from the triple Regge limit of an
inclusive reaction, as shown in Fig.~\ref{fig:TripleReggeLimit}.

In that limit, the double diffraction cross section can be written as \cite{Collins, Goulianos}
$$
\frac{d^2\sigma}{dtdM^2_x}=
\frac{G^{PP, P(t)}_{1 3, 2}}{16\pi^2s_0^2}\left(\frac{s}{s_0}\right)^{2\alpha_P(t)-2}\left(\frac{M^2}{s_0}\right)^{\alpha_P(0)-\alpha_P(t)}.
$$

This approach has two shortcomings. The first one is that it leaves outside the small-$M^2$ resonance region. The second one is connected with the fact that whatever the Pomeron, the (partial) SD cross section overshoots the total one, thus obviously conflicting with unitarity. Various ways of resolving this deficiency
are known from the literature, including the vanishing (decoupling) of the triple Pomeron coupling, but none of them can be considered completely satisfactory.

We instead follow the idea put forward in paper \cite{JL} and developed further in Ref.~\cite{PR}, according to which the Reggeon (here, the Pomeron) is similar to the photon and that the Reggeon-nucleon interaction is similar to deep-inelastic photon-nucleon scattering (DIS), with the replacement $-Q^2=q^2\rightarrow t$ and $s=W^2\rightarrow M^2_x$. There is an obvious difference between the two: while the $C$ parity of the photon is negative, it is positive for the Pomeron. We believe that while the dynamics is essentially invariant under the change of $C$, the difference between the two being accounted for by the proper choice of the parameters. Furthermore, while Jaroszewicz and Landshoff \cite{JL}, in their Pomeron-nucleon DIS structure function (SF) (or $Pp$ total cross section) use the Regge asymptotic limit, we include also the low missing mass, resonance behavior. As is known, gauge invariance requires the DIS SF to vanish as $Q^2$ (here, $t)\ \rightarrow 0$. This property
is built in the SF, see  Refs. \cite{JL, PR}, and it has important consequences for the behavior of the resulting cross sections at low $t$ (see Sec. \ref{sec:Appendix}. Appendix), not shared by the models based on the triple Regge limit, see \cite{Collins, Goulianos}.

It is evident that Regge factorization is essential in both approaches (triple Regge and the present one). It is feasible when Regge singularities are isolated poles. While the pre-LHC data require the inclusion of secondary Reggeons, at the LHC we are in the fortunate situation of a single Pomeron exchange (Pomeron dominance) in the $t$ channel in single and double diffraction (not necessarily so in central diffraction, to be treated elsewhere). Secondary Regge pole exchanges will appear however, in our  dual-Regge treatment of $Pp$ scattering (see below), not to be confused with the the $t$ channel of $pp$.
This new situation makes diffraction at the LHC unique in the sense that for the first time Regge-factorization is directly applicable. We make full use of it.

The paper is organized as follows: Sec. \ref{sec:preLHC} contains a short overview on the pre-LHC measurements of (mainly) single and double diffraction dissociation and their theoretical interpretation. Sec. \ref{sec:Factor} presents simple factorization relations that connecting DD with elastic and SD cross sections. Next, in Sec. \ref{sec:Model} we introduce the dual-Regge model \cite{PR}, the (non-linear) $N^*$ Regge trajectory, the Roper resonance and summarize the main formulae used in our calculations. Fits to the data (figures an tables), presented in Sec. \ref{sec:Results}, are preceded by a brief introduction to our fitting strategy.
In the Appendix (Sec. \ref{sec:Appendix}) we present some results of the main text but with the kinematical factor causing a dramatic turn down of the cross sections as $t$ approaches $0$. Sec. \ref{sec:Conclusions} summarizes our conclusions.

\section{Pre-LHC measurements: CERN ISR and SPS, Fermilab. Generalities} \label{sec:preLHC}
Diffraction dissociation was predicted by theory (for a historical review and relevant references see \cite{Goulianos, Kaidalov, Zotov}) and was intensively studied prior to the LHC on fixed targets (deuteron jets) at the Fermilab by a very successful Soviet-US collaboration \cite{Akimov}  and subsequently at the ISR, SPS and RHIC colliders, see \cite{Goulianos, Tsarev} for more details. Below we briefly summarize the results and conclusions of those studies, mainly of single SD.

1. First measurements concerned low missing masses near the threshold region. Resonance peaks e.g. in the reaction $pp\rightarrow X$, depending on the relevant value of the momentum transfer $t$, were observed beyond the threshold. The origin and properties of theses peaks, still subject of debates, were revealed. The main deficiency of the missing mass method is the lack of any information on the decay properties of the produced particles. The missing information was obtained in bubble chamber experiments and in spark chambers, used also in exclusive channels at the ISR Collider.

Apart from low missing masses, SD was found to persist also to high missing masses. Diffraction (coherence) imposes, however, an upper limit on the highest missing masses $M_x$, roughly as $\xi=M^2_X/s\lesssim 0.05$.

Beyond the resonance region, the smooth $M_x$ dependence of the cross section is approximated by $\frac{d\sigma}{dtdM^2}\sim M^n,$\ \ $n\approx-2$. The value of $n$ is related to the intercept of the trajectories exchanged in the proton-Reggeon scattering, as will be shown in Sec. \ref{subsec:Duality}

2. As expected, SD and DD are peaked in the forward direction, and the slope of the exponential peak of SD was found to be around $8\div12$~GeV$^{-2}$, varying with $s,\ t$ and $M^2$. Near the threshold, the slope is much (about twice) larger than that in elastic scattering, however near $M_x\sim 1.6$~GeV it is already half of that of elastic $pp$. The correlation between the slope parameter and the mass of the excited state is a common feature of SD.

A diffraction minimum around $t\approx 1$~GeV$^2$, similar to that in elastic hadron scattering, is expected also in SD (and DD). There are indications \cite{dip} of such a structure in $pp\rightarrow p(n\pi^+)$ at rather small $|t|$, around $|t|\sim 0.2\div 0.3$~GeV$^2$ at $\sqrt s=53$~GeV, however its origin, fate and affinity to the dip-bump structure in elastic scattering is still a matter of debate \cite{JLL}.

Diffraction is limited both in the missing mass (coherence), $\xi\lesssim 0.05$ and in $t$ ("soft" collisions). There is a transition region in $t$ from "soft" to
"hard" collisions, with a possible dip-bump structure between the two. To be sure, in our analysis we leave outside these interesting but controversial points, concentrating on the "first" cone with clear exponential behavior.

\begin{figure}[hb]
\centering
\includegraphics[width=0.4\linewidth]{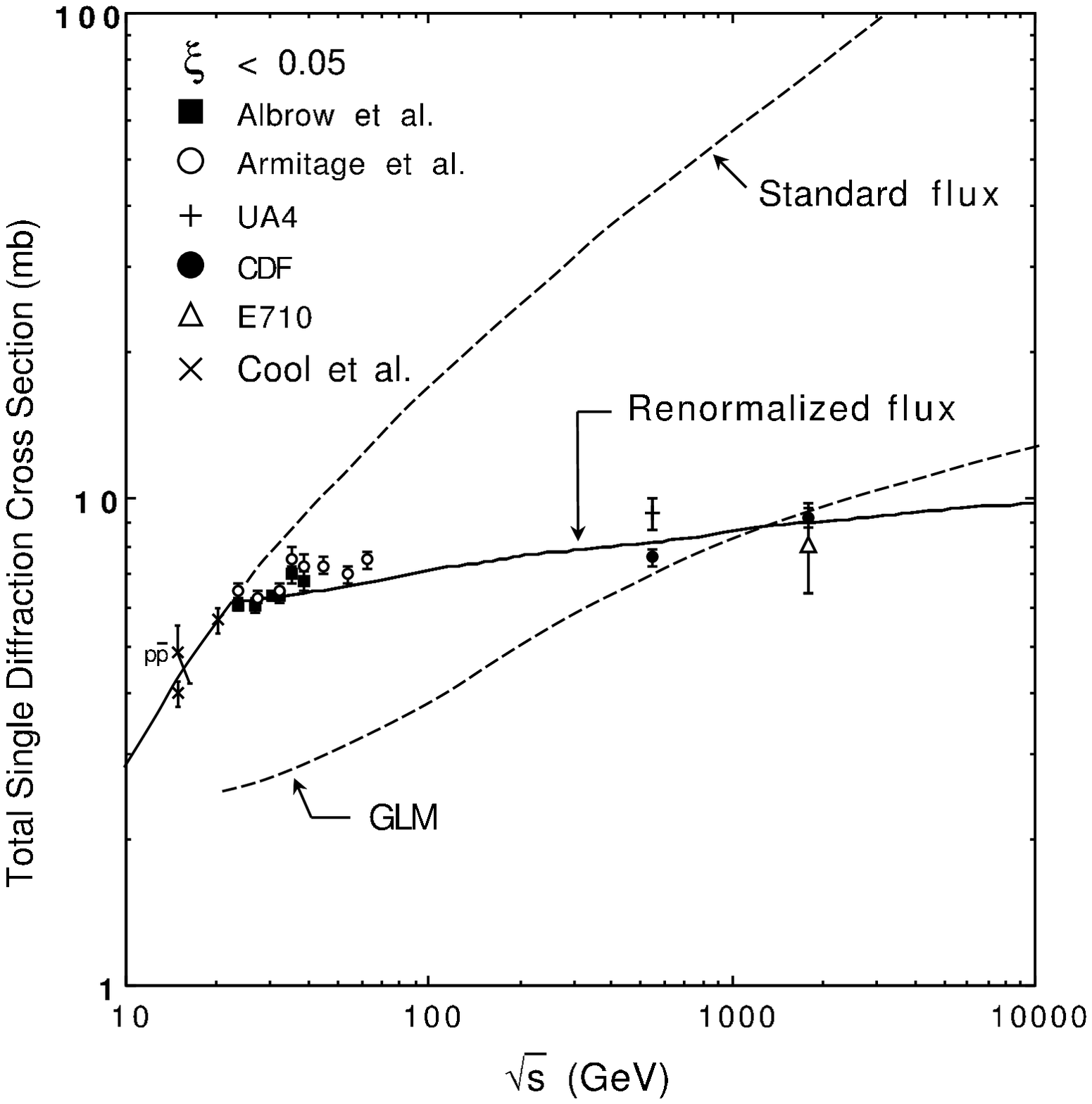}
\caption{Renormalizaton \cite{GoulPlots}}\label{fig:Break}.
\end{figure}

 Before the advent of the LHC, single diffraction dissociation was intensively studied in many different experiments: low energy ISR and SPS CERN experiments, low energy Fermilab (fixed deuteron target) experiments and high energy UA4, UA5, E710 and CDF experiments. All they cover the range $14<\sqrt s<1800$~GeV, $|t|\lesssim2$~GeV$^2$, and missing masses range from the threshold up to $\xi<0.15$. Here $\xi=\frac{M^2}{s}$. The diffraction region is limited up to $\xi<0.15$. At $\xi\sim0.15$ non-diffraction contribution become sizable to diffractive one, and differential cross section become growing with $\xi$.
The main results of these measurements and of their theoretical interpretation can be summarized as follows, for details see, e.g. \cite{Goulianos, Goulianos1}:

1. {\bf Energy dependence.} At energies below 30~GeV the integrated SD cross section rises with $s$ according to the standard prescription of the the Regge-pole theory, however it slows down beyond.
This effect was expected due to the familiar problem related to the violation of unitarity, namely that at high energies, implying the triple Pomeron limit, the DD cross section overshoot the total cross section, $\sigma_{SD}>\sigma_t(s).$ Various means were suggested \cite{triple} to remedy this deficiency, including decoupling (vanishing) of the triple
Pomeron vertex. K. Goulianos instead renormalizes the standard Pomeron flux to meet the data, see \cite{Goulianos1}. Such a "renormalization" produces a break near $\sqrt s$ slowing down the rise of $\sigma_{SD}(s)$ in accord with the CDF data from the Tevatron, as shown in Fig.~\ref{fig:Break}, taken from Ref.~\cite{Goulianos1}.

An alternative approach to resolve this crises (violation of unitarity) is possible within the dipole Pomeron approach Ref.~\cite{JM}.

2. {\bf $t-$ dependence.} SD cross section and the slope $B(s,t,M^2)$ were measured in the range $0.01\lesssim|t|\lesssim2$. The diffraction cone in SD essentially is exponential in $t$; a dip, similar to that in elastic scattering, is likely to appear (somewhere near $t\sim1$~GeV$^2$). In previous work \cite{PR}, the cone of SD differential cross section shows a turn-down towards small-$|t|$ due to the kinematical factor. Experimentally this turn-down was not yet confirmed. We ignore this effect in the main part of this work but will show it in the Appendix.
This tiny effect is located in the kinematical region where Coulomb interaction is sizable, so it may hide the fine structure of the cone. However, as noticed in Ref.~\cite{GJS}, in SD (and DD), Coulomb interaction, at small squared momenta transfers, is suppressed compared to that in elastic diffraction $pp$ scattering, allowing for a better determination of the strongly interacting part of the amplitude (in $pp,$ at small $|t|$ this is possible only indirectly, by means of the Bethe-Heitler interference formula).

\begin{figure}[!ht]
\center{\includegraphics[width=0.4\linewidth]{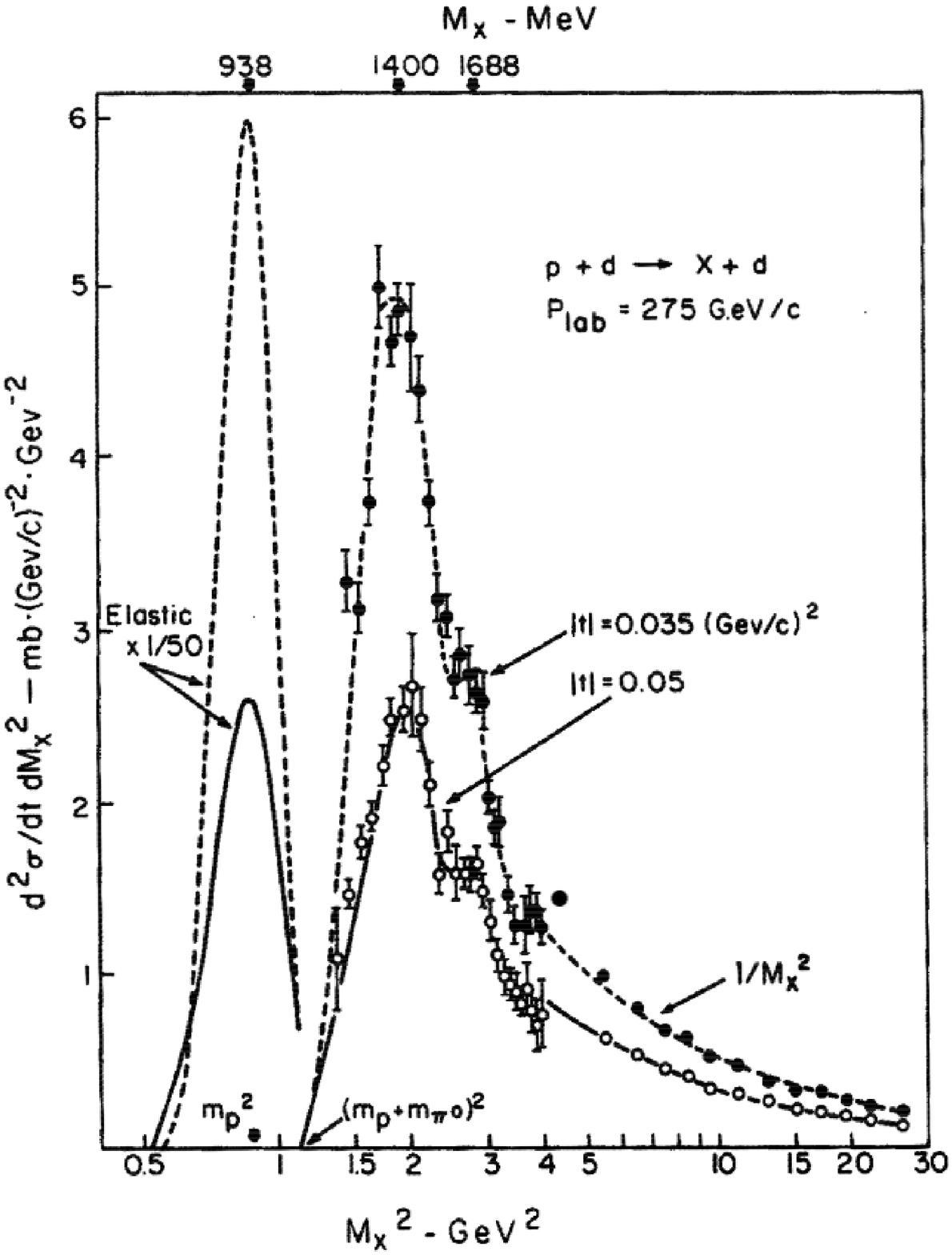}
        \includegraphics[width=0.55\linewidth]{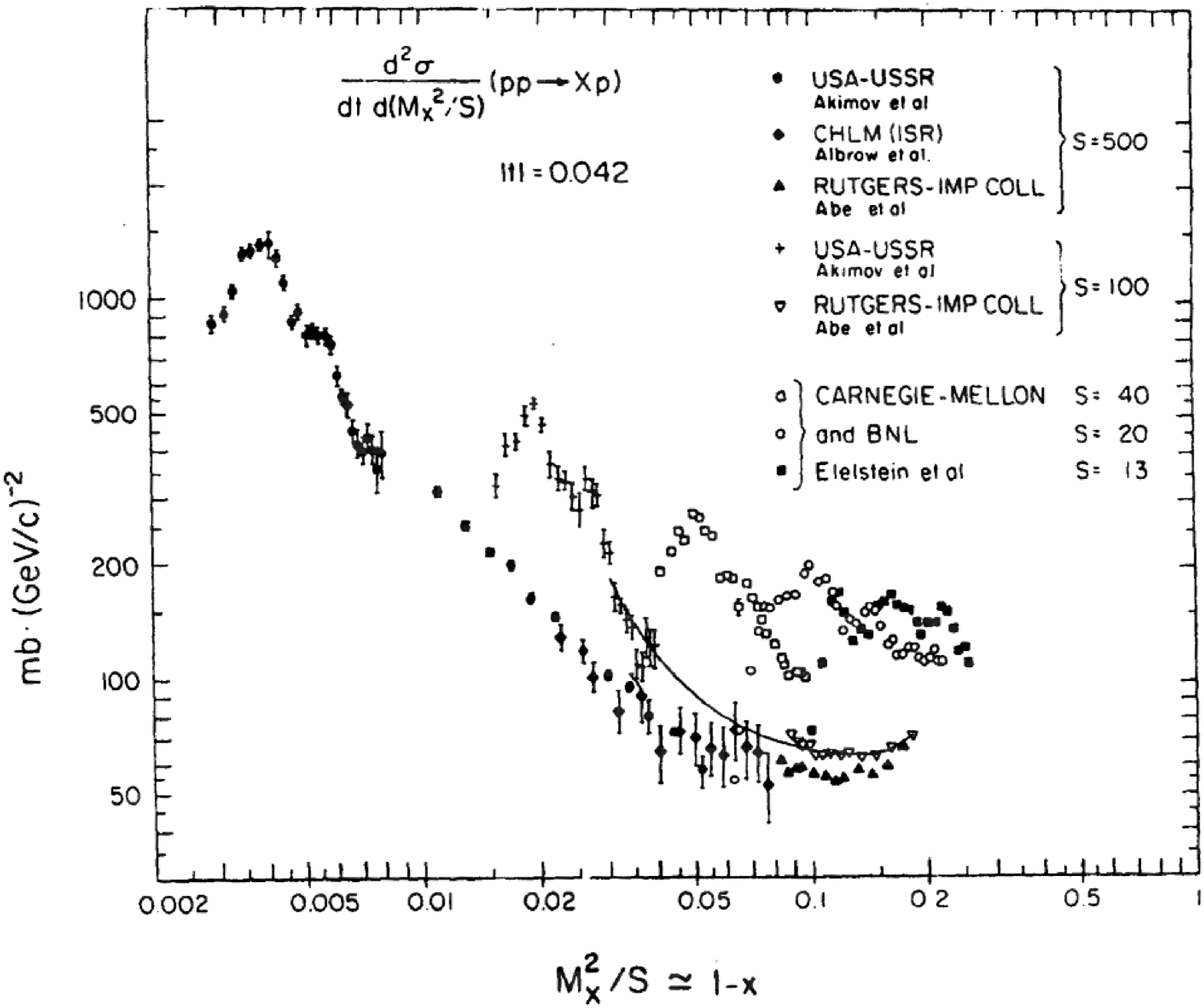}\\
        (a)\hspace{0.45\linewidth}(b)}
\caption{Compilation of low-mass SD data form Fermilab experiments, see: \cite{Goulianos}.}
\label{fig:2SD.Fermilab}
\end{figure}

3. {\bf $M^2$ dependence.} Probably, this is the most delicate issue in the present studies (and diffraction in general). At the ISR, Fermilab and Tevatron, SD was measured in a wide span of the missing mass, starting from the inelastic threshold $M^2_{th}=(m_p+m_{\pi})^2$ up to $\xi<0.05$ (or even to $\xi<0.15$ and higher), Fig.~\ref{fig:2SD.Fermilab}(b).

 As shown in Figs. \ref{fig:2SD.Fermilab}, there is a rich resonance structure in the small $M^2$ region. In most of the papers on the subject, this resonance structure is ignored and replaced by a smooth function $\sim M^{-2}$. Moreover, this simple power-like behavior is extended to the largest available missing masses. In Secs. \ref{sec:Model} and \ref{sec:Results} we question this point on the following reasons: a) The low-$M^2$, resonances introduce strong irregularities in the behavior of the resulting cross sections. LHC measurements, are able to probe low-$M^2$ region, and will be sensitive to these structures. b) The large-$M$ behavior of the amplitude (cross sections) is another delicate point. Essentially, it is determined by the proton-Pomeron ($pP$) total cross section, proportional to the $pP$ structure function, discussed in details in Sec. \ref{sec:Model}.
By duality, the averaged contribution from resonances sums up to produce high missing mass Regge behavior $~(M^2)^{-n},$ where $n$ is related to the intercept of the exchanged Reggeon and may be close (but not necessarily equal) to the above-mentioned empirical value $\sim 1$.


In our calculations of the large-$M$ behavior (Secs. \ref{sec:Model} and \ref{sec:Results}) we shall refer also to the ISR-Tevatron-CDF data, remembering that for low energies ($\lesssim500$~GeV), non-leading Regge exchanges are also important \cite{JLL}.

\section{Simple (and approximate) factorization relations} \label{sec:Factor}
With the advent of the LHC, diffraction, elastic and inelastic scattering entered a new area, where it can be seen uncontaminated by non-diffraction
events. In terms of the Regge-pole theory this means, that the scattering amplitude is completely determined by a Pomeron exchange, and in a
simple-pole approximation, Regge factorization holds and it is of practical use! Remind that the Pomeron is not necessarily a simple pole:
perturbative QCD suggests that the Pomeron is made of an infinite number of poles (useless in practice), and the unitarity condition requires corrections
to the simple pole, whose calculation is far from unique. Instead a simple Pomeron pole approximation \cite{DL} is efficient in describing
a variety of diffraction phenomena.

\begin{figure}[!ht]
\includegraphics[width=0.8\textwidth]{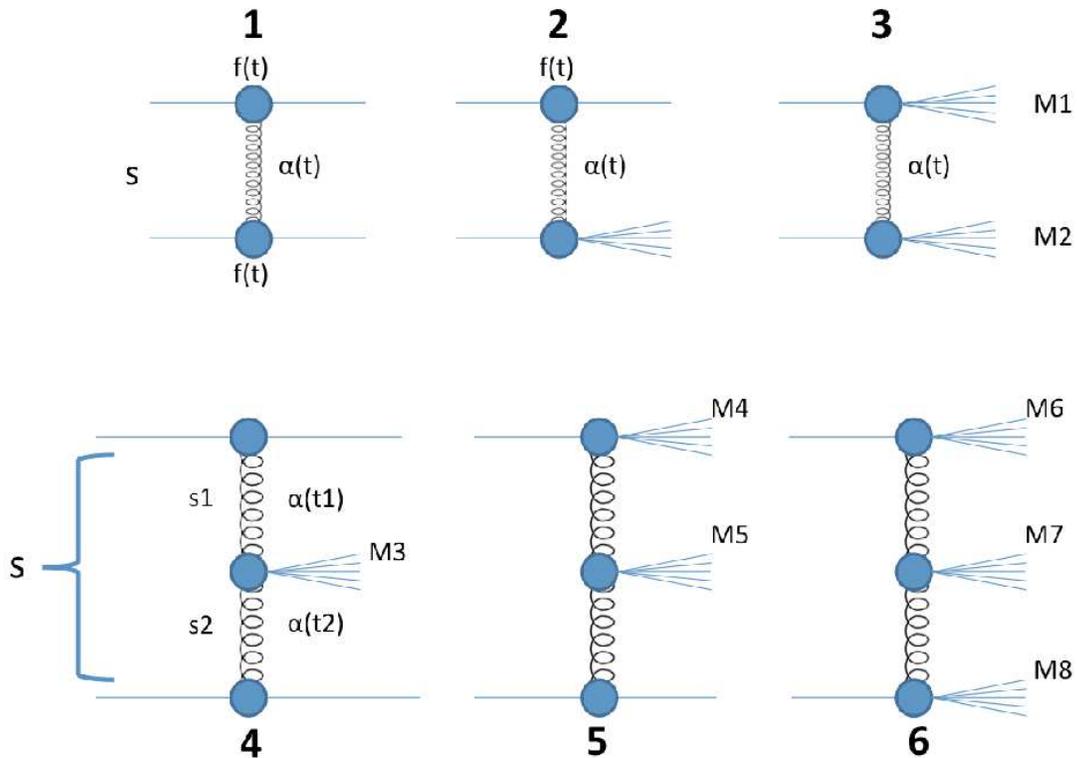}
\caption{Diagrams for elastic scattering and diffraction dissociation (single, double and central).} \label{fig:FeynmanDiagrams}
\end{figure}

The D-L elastic scattering amplitude is simply \cite{DL}
\begin{equation}\label{DL}
A(s,t)=\xi(t)\beta(t)^2(s/s_0)^{\alpha_P(t)-1}+A_R(s,t),
\end{equation}
where $\xi(t)$ is the signature factor, and $\alpha(t)$ is the (linear) Pomeron trajectory. The signature factor can be
written as $\xi(t)=e^{-i\pi/2}$, however it is irrelevant here, since below we shall use only cross sections (squared modules of the amplitude), where it reduces to unity. The residue is chosen to be a simple exponential, $\beta(t)=e^{b_Pt}$.
"Minus one" in the propagator term $(s/s_0)^{\alpha_P(t)-1}$ of (\ref{DL}) correspond for normalization $\sigma_T(s)=Im A(s,t=0)$. The scale parameter $s_0$ is not fixed by the Regge-pole theory: it can be fitted do the data or fixed to a "plausible" value of a hadronic mass, or to the inverse "string tension" (inverse of the Pomeron slope), $s_0=1/\alpha'$. The second term in Eq.~(\ref{DL}), corresponding to sub-leading Reggeons, has the same functional form as the first one (that of the Pomeron), just the values of the parameters differ.
We ignore this term for reason mentioned above.

Fig.~\ref{fig:FeynmanDiagrams} shows the simplest configurations of Regge-pole diagrams for elastic, single- and double diffraction dissociation, as well as central diffraction dissociation (CD). In this paper we consider only SD and DD.

Factorization of the Regge residue $\beta(t)$ and the "propagator" $(s/s_0)^{\alpha_P(t)-1}$ is a basic property of the theory. As mentioned, at the LHC for the first time,
we have the opportunity to test directly Regge-factorization  in diffraction, since the scattering amplitude here is dominated by a simple Pomeron-pole exchange, identical in elastic and inelastic diffraction.
Simple factorization relations between elastic ($\frac{d\sigma_{el}}{dt}$), single ($\frac{d\sigma_{el}}{dt})$ and double ($\frac{d^3\sigma_{DD}}{dtdM_1^2dM_2^2})$ DD are known from the literature \cite{Goulianos}.
Really, by writing the scattering amplitude as product of the vertices, elastic $f$ and inelastic $F$, multiplied by the (universal) propagator (Pomeron exchange), $f^2s^{\alpha} \ \ fFs^{\alpha},\ \ F^2s^{\alpha}$ for elastic scattering, SD and DD, respectively, one gets
\begin{equation}\label{factor1}
\frac{d^3\sigma_{DD}}{dtdM_1^2dM_2^2}=\frac{d^2\sigma_{SD1}}{dtdM_1^2}\frac{d^2\sigma_{SD2}}{dtdM_2^2}/\frac{d\sigma_{el}}{dt}.
\end{equation}

Assuming $e^{Bt}$ as a $t-$dependence for both SD and elastic scattering, integration over $t$ yields:
\begin{equation}\label{factor2}
\frac{d^3\sigma_{DD}}{dM_1^2dM_2^2}=k\frac{d^2\sigma_{SD1}}{dM_1^2}\frac{d^2\sigma_{SD2}}{dM_2^2}/\sigma_{el}.
\end{equation}
where $k=r^2/(2r-1),\ \ r=B_{SD}/B_{el}$.

For $pp$ interactions at the ISR, $r=2/3$ and hence $k=4/3$. Taking the value $r=2/3$, consistent with the experimental results at Fermilab and ISR, one obtains
$\sigma_{DD}=\frac{4\sigma_{SD}}{3\sigma_{el}},$. Notice that for $r=1/2,\ \ k \rightarrow\infty.$  Thus $k$ is very sensitive to the ratio $r$, which shows that direct measurements of the slopes at the LHC are important. Interestingly, relation (\ref{factor2}) can be used in different ways, e.g. to cross check any among the four inputs.

To summarize this discussion, we emphasize the important role of the ratio between the inelastic and elastic slope, which at the LHC is close to its
critical value $B_{SD}/B_{el}=0.5$ (it cannot go below!), which means a very sensitive correlation between these two quantities. The right balance may require a
correlated study of the two by keeping the ratio above $0.5$. This constrain may guide future experiments on elastic and inelastic diffraction.
For detailed details see also \cite{Chew:1974zy,Chew:1974vu}.

\section{Model for single and double diffraction dissociation}\label{sec:Model}
The model relies on the following premises:

1.  {\bf Regge factorization} is feasible since, as stressed repeatedly, at the LHC energies in the region of $|t|<1$~GeV$^2$, which is typical for diffraction, the contribution from secondary Reggeons is negligible, and, for a single Pomeron term, factorization (\ref{factor1}) is exact. Due to factorization, the relevant expressions for the cross sections (elastic, SD, DD) have simple forms (\ref{elastic}), (\ref{SD}), (\ref{DD}). Such relations are known from the literature, see e.g. \cite{Goulianos, Chew:1974zy,Chew:1974vu} and references therein.

2. {\bf The inelastic $pPX$ vertex} receives special care.

Having justified and accepted the factorized form of the scattering amplitude,
the main object of our study is now the inelastic proton-Pomeron vertex or transition amplitude.
As argued in Refs. \cite{JL,PR}, it can be treated as the proton structure function (SF), probed by the Pomeron, and proportional to the
Pomeron-proton total cross section, $\sigma_T^{Pp}(M^2_x, t),$ with the norm $\sigma_T^{Pp}(M^2_x, t)={\cal I} m A(M^2,t),$
in analogy with the proton SF probed by a photon (in $ep$ scattering e.g. at HERA or JLab).
$$\nu W_2(M_x^2,t)=F_2(x,t)=\frac{4(-t)(1-x)^2}{\alpha(M_x^2-m_p^2)(1+4m_p^2x^2/(-t))^{3/2}}{\cal I}m A(M^2,t),$$
where $\alpha$ is the fine structure constant, $\nu=\frac{M_x^2-m_p^2-t}{2m_p}$, and $x=\frac{-t}{2m_p\nu}$ is the Bjorken variable.

The only difference is that the Pomeron's (positive) $C$ parity is opposite to that of the photon. This difference is evident in the values of the parameters but is unlikely to affect the functional form of the SF itself, for which we choose its high-$M_x^2$ (low Bjorken $x$) behavior. Notice that the the total energy in this subprocess, the analogy of $s=W^2$ in DIS, here is $M^2_x$ and $t$ here replaces $q^2=-Q^2$ of DIS.
Notice that gauge invariance requires that the SF vanishes towards $Q^2\rightarrow 0$ (here, $t$), resulting in the dramatic vanishing of the SD and DD differential cross section towards $t=0.$  How fast does the SF (and relevant cross sections) recover from $t=0$ a priori is not known.

\begin{figure}[!ht]
 \centering
 \includegraphics[width=0.24\linewidth,bb=0 0 480 480,clip]{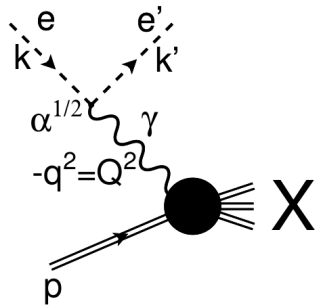}
 \caption{ Virtual photon + proton $\rightarrow M_x^2$ transition.}
 \label{fig:gamma_p}
\end{figure}

Furthermore, according to the ideas of two-component duality, see e.g. \cite{JPsi}, the cross sections of any process, including that of $pP\rightarrow X,$ is a sum of a non-diffraction component, in which resonances sum up in high-energy (here: mass $M^2$ plays the role of energy $s$) Regge exchanges and the smooth background (below the resonances), dual to the Pomeron exchange. The dual properties of diffraction dissociation can be quantified also by finite mass sum rules, see \cite{Goulianos}. In short: the high-mass behavior of the $pP\rightarrow X$ cross section is a sum of a decreasing term going like $\sim \frac{1}{M^m}$,$m\approx 2$ and a "Pomeron exchange" increasing slowly with mass. All this has little affect on the low-mass behavior at the LHC, however normalization implies calculation of cross sections integrated over all physical values of $M^2$, i.e. until $M^2<0.05s$.

3. {\bf The background} is a delicate issue. In the reactions (SD, DD) under consideration there are two sources of the background. The first is that related to the $t$ channel exchange in Fig \ref{fig:2SD.Fermilab}(b) and it can be accounted for by rescaling the parameter $s_0$ in the denominator of the Pomeron propagator. In any case, at high energies, those of the LHC, this background is included automatically in the Pomeron. The second component of background comes from the subprocesses $pP\rightarrow X$. Its high-mass behavior is not known experimentally and it can be only conjecture on the bases of the known energy dependence of the typical meson-baryon processes appended by the ideas of duality. The conclusion is that the $Pp$ total cross section at high energies (here: missing masses $M$) has two components:
 a decreasing one, dual to direct-channel resonances and going as $\sigma_{tot}^{Pp}\sim \sum_R(s')^{\alpha_R(0)-1}= \sum_R\left(M^2\right)^{\alpha_R(0)-1},$ where $R$ are non-leading Reggeons, and a slowly rising Pomeron term producing $\sim M^{2\cdot0.08}$ \cite{DL}.

\subsection{Duality}\label{subsec:Duality}
\begin{figure}[htb]
 \includegraphics[width=.9\textwidth]{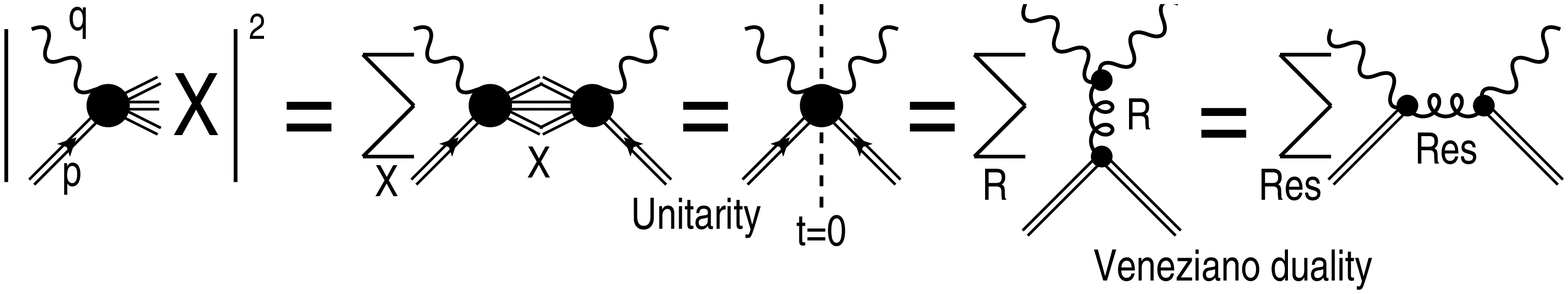}
\caption{Connection, through unitarity (generalized optical
theorem) and Veneziano-duality, between the inelastic form factor
and sum of direct-channel resonances.} \label{fig:Regge-dual}
\end{figure}

Any meson-baryon total cross section (or scattering amplitude) is a sum of two contributions: diffractive and non-diffractive. By the concept of two-component duality \cite{H_R},
the diffractive component, the smooth background at low energies (here: missing masses) is dual to a Pomeron $t$-channel exchange at high energies, while the non-diffractive
component contains direct channel resonances, dual to high-energy $t-$ channel (sub-leading) Reggeon exchanges, as shown in Fig.~\ref{fig:Regge-dual}.

According to our present knowledge about two-body hadronic reactions, two distinct classes of reaction mechanisms exist.


The first one includes the formation of resonances in the $s-$channel
and the exchange of particles, resonances, or Regge
trajectories in the $t-$channel. The low-energy resonance
behavior and the high-energy Regge asymptotics are related by
duality, which at Born level, or,
alternatively, for tree diagrams, mathematically can be formalized
in the Veneziano model, which is a combination of Euler
Beta-functions \cite{Veneziano}.


The second class of mechanisms does not exhibit resonances at low
energies and its high-energy behavior is governed by the exchange
of a vacuum Regge trajectory, the Pomeron, with an intercept equal
to or slightly greater than one. Harari and Rosner \cite{H_R}
hypothesized that the low-energy non-resonating background is dual
to the high-energy Pomeron exchange, or diffraction. In other
words, the low energy background should extrapolate to high-energy
diffraction in the same way as the sum of narrow resonances sum up
to produce Regge behaviour. However, contrary to the case of
narrow resonances, the Veneziano amplitude, by construction,
cannot be applied to (infinitely) broad resonances. This becomes
possible in a generalization of narrow resonance dual models
called dual amplitudes with Mandelstam analyticity (DAMA),
allowing for (infinitely) broad resonances, see \cite{DAMA} and references therein, or
the background.

In the resonance region, roughly $1\lesssim M\lesssim 4$~GeV, the non-diffractive component of the amplitude is adequately described by a "reggeized Breit-Wigner" term  Eq.~(\ref{total}), following from the low-energy decomposition of a dual amplitude, with a direct-channel meson-baryon (Pomeron-proton, in our case) trajectory (see Fig.~\ref{fig:n1} in Sec. \ref{Sec:trajectory}), with relevant nucleon resonances lying on it, appended by a Roper resonance (see Eq.~(\ref{Roper_eq})).


 \begin{table}[!hb]
 \begin{tabular}{|c|c|c|}
   \hline
   ${\cal I}m A(a+b\rightarrow c+d)=$& $R$& Pomeron \\
   \hline
   $s-$channel & $\sum A_{Res}$  & Non-resonant background \\
   \hline
   $t-$channel & $\sum A_{Regge}$ & Pomeron $(I=S=B=0;\ C=+1)$ \\
   \hline
   High energy dependence & $s^{\alpha-1},\ \alpha<1$ & $s^{\alpha-1},\ \alpha\geq 1$ \\
   \hline
 \end{tabular}
 \caption{Two-component duality}
 \end{table}

By duality, a proper sum of direct channel resonances produces smooth Regge behavior, and, to avoid "double counting", one should not add the two. Actually, this is true only for an infinite number of resonance poles. For technical reasons, we include only a finite number of resonance poles, moreover, apart from the "regular" contribution of the nucleon resonances, lying on the $N^*$ trajectory, the Roper resonance is also included (see Sec. \ref{sec:Roper}). The spectroscopic status of this resonance is disputable. It has no place on the known baryonic trajectories, but its hight exceeds that of the neighboring, next most important $N^*(1680)$. In view of the "truncated" series of resonance poles, we do not expect that it will reproduce correctly the high-energy Regge behavior, therefore we add it to the total cross section in the form of an "effective" Regge pole contribution.

The second, diffractive component, is essentially the contribution from a Pomeron pole exchange
$\sigma^{Pp}_T\sim (M^2)^{\alpha(0)-1}$, with $\alpha(0)\approx 1.08$ \cite{DL},
and as shown by Donnachie and Landshoff \cite{DL} this term also can give some contribution to the low-energy (here, missing mass) flat background.

To summarize this discussion, Regge-pole exchanges take place at two distinct parts of the diagrams shown in Fig.~~\ref{fig:FeynmanDiagrams}: in the $t$ channel, where, at the LHC, only the Pomeron contributes, and in the inelastic form factor, sub-diagram shown in Fig.~~\ref{fig:gamma_p}, where, depending of the value of the missing mass, both the Pomeron and Reggeons are equally important. At low missing masses, the direct-channel proton trajectory $N^*$, dominates, replaced by an effective Reggeon exchange at high masses, appended by a Pomeron. The Roper resonance (with its controversial status) in the direct channel stay apart. Although we concentrate on the low missing mass region, the behavior of the cross sections at high masses are important in the calculation of the cross sections integrated in $M^2$. We remind that we impose a limit on diffraction events to be about $\xi<0.05$ or $M<200$~GeV (as it used at some experiments).

\subsection{Resonances in the $Pp$ system; the $N^*$ trajectory}\label{Sec:trajectory}
The Pp total cross section at low missing masses is dominated by nucleon resonances. In the dual-Regge approach \cite{PR}, the relevant cross section is a "Breit-Wigner" sum Eq.~(\ref{total}), in which the direct-channel trajectory is that of $N^*$. An explicit model of the nonlinear $N^*$ trajectory was elaborated in paper \cite{Paccanoni} and used in \cite{PR}.

\begin{equation} \label{total}
\sigma^{Pp}_T(M_x^2,t)={\cal I}m\, A(M_x^2,t)={\cal I} m\left(\sum_{n=0,1,...}\frac{a f(t)^{2(n+1)}}{2n+0.5-\alpha_{N^*}(M_x^2)}\right).
\end{equation}
%

\begin{figure}[htb]
\includegraphics[width=.49\textwidth,bb= 10 140 540 660]{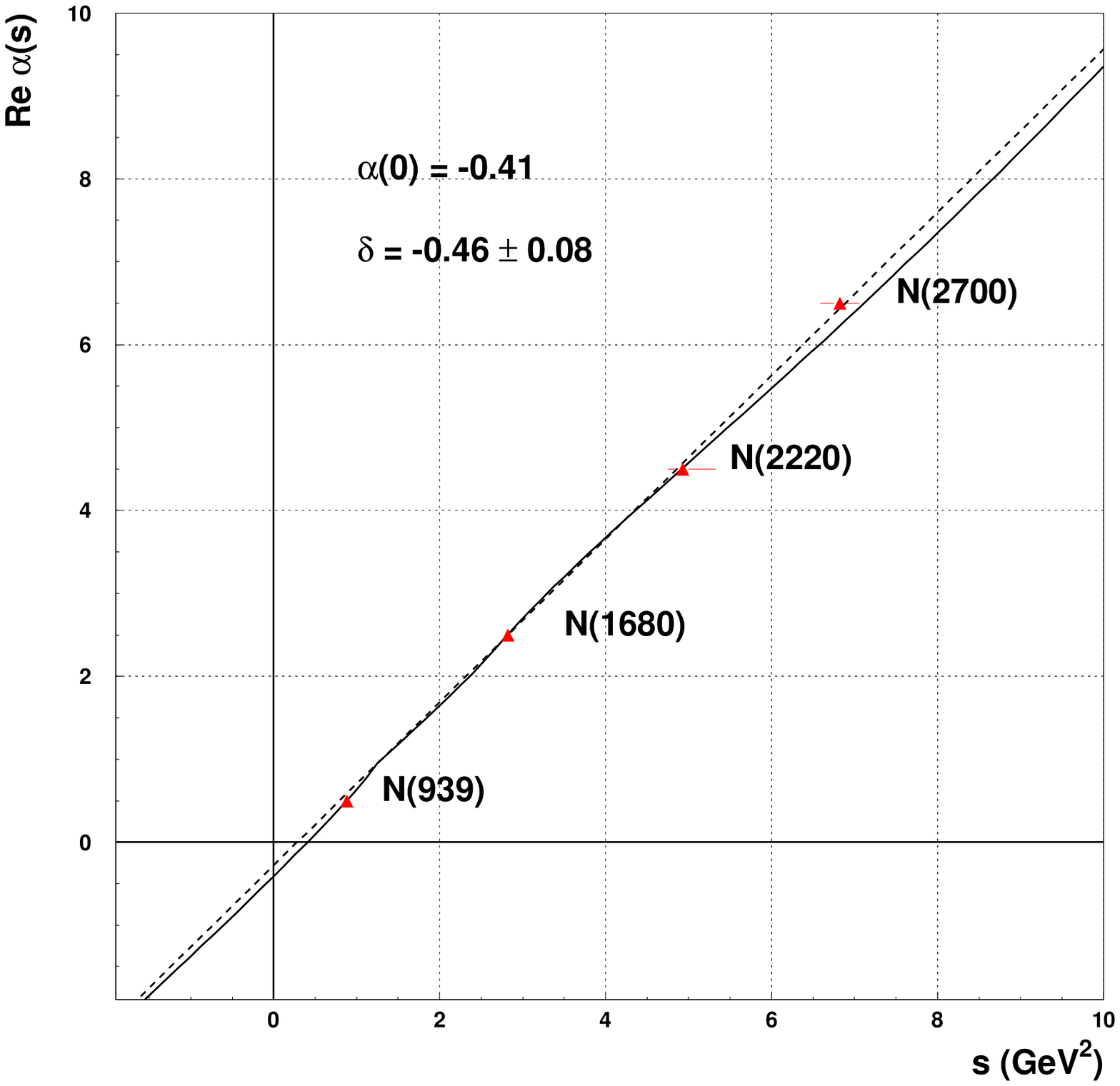}
\includegraphics[width=.49\textwidth,bb= 10 140 540 660]{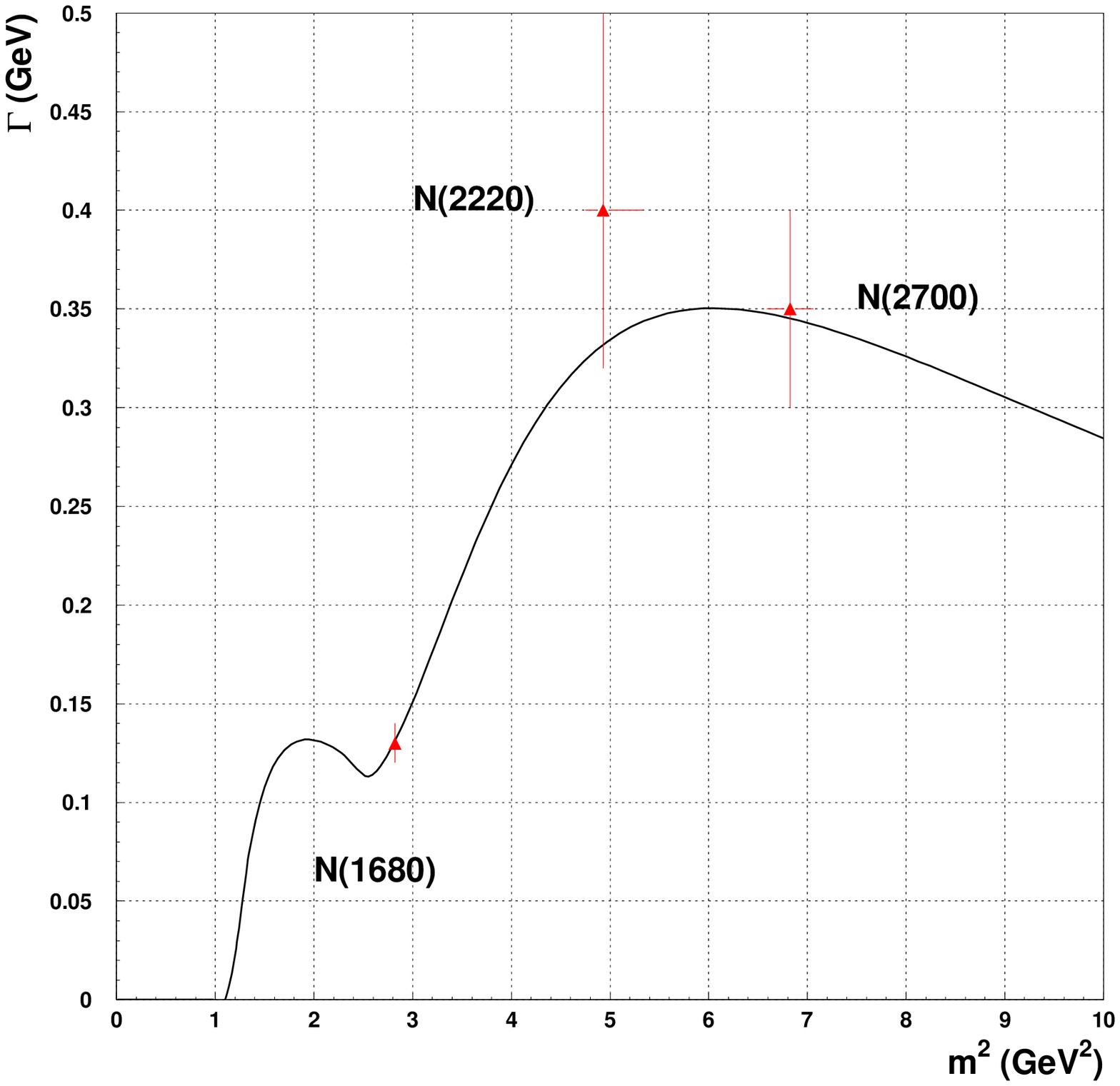}
\parbox[t]{8.5cm}{\caption{The real part of the proton Regge trajectory. The dashed line
corresponds to the result of a linear fit, the solid line is the
fit from \cite{Paccanoni}. \label{fig:n1}}}
\hfill~\parbox[t]{8.cm}{\caption{ The widths of the resonances,
$\Gamma=\frac{{\cal I}m\,\alpha(M^2)}{M{\cal R}e\,\alpha'(M^2)}~,$
appearing on the proton trajectory, calculated and fitted in
\cite{Paccanoni}. \label{fig:n1width}}}
\end{figure}

The Pomeron-proton channel, $Pp\rightarrow M_X^2$  couples to the
proton trajectory, with the $I(J^P)$ resonances: $1/2(5/2^+),\
F_{15},\ m=1680$~MeV, $\Gamma=130$~MeV; $1/2(9/2^+),\ H_{19},\
m=2200$~MeV, $\Gamma=400$~MeV; and $1/2(13/2^+),\ K_{1,13},\
m=2700$~MeV, $\Gamma=350$~MeV. The status of the first two is
firmly established \cite{particles}, while the third one,
$N^*(2700),$ is less certain, with its width varying between
$350\pm 50$ and $900\pm 150$~MeV . Still, with the
stable proton included, we have a fairly rich trajectory,
$\alpha(M^2)$, whose real part is shown in Fig.~\ref{fig:n1}.

Despite the seemingly linear form of the trajectory, it is not
that: the trajectory must contain an imaginary part corresponding
to the finite widths of the resonances on it. The non-trivial
problem of combining the nearly linear and real function with its
imaginary part was solved in Ref.~\cite{Paccanoni} by means of
dispersion relations.

We use  the explicit form of the trajectory derived in Ref.~
\cite{Paccanoni}, ensuring  correct behaviour of both its real and
imaginary parts. The imaginary part of the trajectory can be
written in the following way:
\begin{equation}
{\cal I}m\, \alpha(s)=s^{\delta} \sum_n c_n
\left(\frac{s-s_n}{s}\right)^{\lambda_n} \cdot \theta(s-s_n)\,,
\label{b2}
\end{equation}
where $\lambda_n={\cal R}e\ \alpha(s_n)$. Eq.~~(\ref{b2}) has the
correct threshold behaviour, while analyticity requires that
$\delta <1$. The boundedness of $\alpha(s)$ for $s \to \infty$
follows from the condition that the amplitude, in the Regge form,
should have no essential singularity at infinity in the cut plane.

The real part of the proton trajectory is given by
\begin{equation}
{\cal R}e\,\alpha(s)=\alpha(0)+\frac{s}{\pi}\sum_n c_n {\cal
A}_n(s)~, \label{b3}
\end{equation}
where
\begin{eqnarray*}
& & {\cal A}_n(s)=
\frac{\Gamma(1-\delta)\Gamma(\lambda_n+1)}{\Gamma(\lambda_n-\delta+2)
s_n^{1-\delta}}{}_2F_1\left(1,1-\delta;\lambda_n-\delta+2;\frac{s}{s_n}\right)\theta(s_n-s)+\\
& & \left\{ \pi s^{\delta-1}\left(
\frac{s-s_n}{s}\right)^{\lambda_n} \cot[\pi(1-\delta)]- \right. \\
& & \left.
\frac{\Gamma(-\delta)\Gamma(\lambda_n+1)s_n^{\delta}}{s\Gamma(\lambda_n-\delta+1)
}{}_2F_1\left(\delta-\lambda_n,1;\delta+1;\frac{s_n}{s}\right)
\vphantom{\left( \frac{s-s_n}{s}\right)^{\lambda_n}} \right \}
\theta(s-s_n)~.
\end{eqnarray*}

The proton trajectory, also called $N^+$
trajectory \cite{Paccanoni}, contains the baryons N(939)
$\frac{1}{2}^+$, N(1680) $\frac{5}{2}^+$, N(2220) $\frac{9}{2}^+$
and N(2700) $\frac{13}{2}^+$ \cite{particles}. In the fit, the
input data are the masses and widths of the resonances. The
quantities to be determined are the parameters $c_n$, $\delta$ and
the thresholds $s_n$. Following \cite{Paccanoni} we set $n=1,2,x$
and $s_1=(m_{\pi}+m_N)^2=1.16$~GeV$^2$, $s_2$ = 2.44~GeV$^2$ and
$s_x$ = 11.7~GeV$^2$.

Other parameters of the trajectory, obtained in the fit, are
summarized below: $\alpha(0)=-0.41$, $\delta = -0.46 \pm 0.07$,
$c_1=0.51\pm 0.08$, $c_2=4.0\pm 0.8$ and $c_x=(4.6 \pm 1.7)\cdot
10^{3}$. Taking the central values of these parameters we obtain
the following values for the $\lambda$'s: $\lambda_1=0.846$,
$\lambda_2=2.082$,  $\lambda_x=11.177$.



The elastic contribution, is separated from SD and DD by a gap extending from the proton mass $m_p$ to the first threshold, at $m_p+m_{\pi},$
and it should be treated separately.

Thus, we obtain:
\begin{equation}\label{ImA_fin}
{\cal I}m\,  A(M_X^2,t)=a \sum_{n=1,3}[f(t)]^{2(n+1)}
\frac{{\cal I}m\, \alpha(M^2_X)}{(2n+0.5-Re\, \alpha(M_X^2))^2+
({\cal I}m\, \alpha(M_X^2))^2}\,.
\end{equation}

\subsection{The Roper resonance}
\label{sec:Roper}

Apart from the well established protonic trajectory
with a sequence of four particles, there is a prominent single resonance
$I=1/2,\ \  J=1/2^+$ with mass $1440$~MeV, known as the Roper
resonance \cite{particles}. It is wide, the width being nearly one
fourth of its mass, its spectroscopic status being disputable.
There is no room for the Roper resonance on the proton trajectory
of Sec. \ref{Sec:trajectory}, although it could still be a member of protons
daughter trajectory. Waiting for a future better understanding of Roper's status, here we
present the contribution to SD cross section of a single Roper
resonance, calculated from a simple
Breit-Wigner formula:
\begin{equation}\label{Roper_eq}
{\cal I} m A_{Roper}(M_x^2,t)=
b \frac{f^2(t) M_{Roper} \Gamma_{Roper}/2 }{(M^2_x - M^2_{Roper})^2 + (\Gamma_{Roper}/2)^2},
\end{equation}
where $M_{Roper} = 1440$~MeV, $\Gamma_{Roper} = 325$~MeV,  and $c$ is another normalization parameter.

\subsection{Compilation of the basic formulae}
This subsection contains a compilation of the main formulae used in the calculations and fits to the data.

The elastic cross section is:
  \begin{equation}\label{elastic}\frac{d\sigma_{el}}{dt}=A_{el}{F_p}^4(t)\left(\frac{s}{s_0}\right)^{2(\alpha(t)-1)}.\end{equation}
The single diffraction (SD) dissociation cross section is:
  \begin{equation}\label{SD}2\cdot\frac{d^2\sigma_{SD}}{dtdM_x^2}={F_p}^2(t){F_{inel}}^2(t,M_x^2) \left(\frac{s}{M_x^2}\right)^{2(\alpha(t)-1)}.\end{equation}
Double diffraction (DD) dissociation cross section:
  \begin{equation}\label{DD}\frac{d^3\sigma_{DD}}{dtdM_1^2dM_2^2}=N_{DD}{F_{inel}}^2(t,M_1^2){F_{inel}}^2(t,M_2^2)\left(\frac{ss_0}{M_1^2M_2^2}\right)^{2(\alpha(t)-1)}.\end{equation}
with the norm $N_{DD}=\frac{1}{4A_{el}},$
with the inelastic vertex:
  \begin{equation}\label{eq:inelVertex}
   {F_{inel}}^2(t,M_x^2)=A_{res}\frac{1}{M_x^4}\sigma_T^{Pp}(M_i^2,t)+C_{bg}\sigma_{Bg},
  \end{equation}
where the Pomeron-proton total cross section is the sum $N^*$ resonances (Eq.~(\ref{ImA_fin})) and the Roper resonance (Eq.~(\ref{Roper_eq})), with a relevant norm factor $R$ (we remove the $t$ dependent $f_{res}(t)$ out of the sum):
  \begin{equation}\label{eq:sigma_tot}
    \sigma_T^{Pp}(M_x^2,t)=
    R\frac{[f_{res}(t)]^{2} \cdot M_{Roper}\left(\frac{\Gamma_{Roper}}{2}\right)}
    {{\left(M_x^2-M_{Roper}^2\right)}^2+{\left(\frac{\Gamma_{Roper}}{2}\right)}^2}
    +[f_{res}(t)]^{4}\sum_{n=1,3} \frac{{\cal I} m\,\alpha(M^2_x)}{(2n+0.5-{\cal R} e\, \alpha(M_x^2))^2+({\cal I} m\,\alpha(M^2_x))^2},
  \end{equation}
and the background corresponding to non-resonance contributions:
  \begin{equation}\label{eq:sigma_bg}
   \sigma_{Bg}=\frac{f_{bg}(t)}{\frac{1}{{\left(M_x^2-(m_p+m_{\pi})^2\right)}^\varsigma}+(M_x^2)^{\eta}},
  \end{equation}
  {\it{\bf NB:} In Eq.~(\ref{eq:sigma_bg}) $M_x$, $m_p$ and $m_p$ are in $[$GeV$]$.}

 The Pomeron trajectory is \cite{JLL}:
  $$\alpha(t)=1.075+0.34t,$$
and the  $t-$dependente elastic and inelastic form factors are:
  $$F_p(t)=e^{b_{el}t}, \qquad f_{res}(t)=e^{b_{res}t}, \qquad f_{bg}(t)=e^{b_{bg}t}.$$
  The slope of the cone is defined as:
  \begin{equation}\label{eq:B}
    B=\frac{d}{dt}\ln\frac{d\sigma}{dt},
  \end{equation}
where $\frac{d\sigma}{dt}$ stands for $\frac{d\sigma_{el}}{dt},$  $\frac{d\sigma_{SD}}{dt},$ or $\frac{d\sigma_{DD}}{dt}$,  defined by Eq.~(\ref{elastic}), (\ref{SD}) and  (\ref{DD}), respectively.

  The local slope $B_M$ at fixed $M_x^2$ is defined in the same way:
  \begin{equation}\label{eq:B_M}
    B_{M}=\frac{d}{dt}\ln\frac{d^2\sigma}{dtdM_x^2}
  \end{equation}
  {\it{\bf NB:} $\frac{d\sigma}{dt}$ in Eq.~(\ref{eq:B}) is in units of $[mb\cdot$GeV$^{-2}]$, and $\frac{d^2\sigma}{dtdM_x^2}$ in Eq.~(\ref{eq:B_M})  in units of $[mb\cdot$GeV$^{-4}]$.}

  The integrated cross sections are calculated as:
  \begin{equation}\label{eq:dcsdt_SD}
    \frac{d\sigma_{SD}}{dt}=\int_{M^2_1}^{M^2_2}\frac{d^2\sigma_{SD}}{dtdM_x^2}dM_x^2
  \end{equation}
  for the case of SD and:
  \begin{equation}\label{eq:dcsdt_DD}
    \frac{d\sigma_{DD}}{dt}= \int\int_{f(M^2_{x_1},M^2_{x_2})}\frac{d^3\sigma_{SD}}{ dtdM_{x_1}^2 dM_{x_2}^2 }dM_{x_1}^2dM_{x_2}^2
  \end{equation}
  for the case of DD.\\
  We also calculate the fully integrated cross sections:
  \begin{equation}
  \sigma_{SD}=\int_{0}^{1}dt\int_{M^2_{th}}^{0.05s}dM_{x}^2 \frac{d^2\sigma_{SD}}{dtdM_{x}^2},
  \end{equation}
  \begin{equation}
   \sigma_{DD}=\int_{0}^{1}dt\int\int_{\Delta\eta>3}dM_{x_1}^2dM_{x_2}^2 \frac{d^3\sigma_{DD}}{dtdM_{x_1}^2dM_{x_2}^2}
  \end{equation}
  and
  \begin{equation}
  \frac{d^2\sigma_{DD}}{dM_{x_1}^2dM_{x_2}^2}= \int_{0}^{1}\frac{d^3\sigma_{DD}}{dtdM_{x_1}^2dM_{x_2}^2}dt.
  \end{equation}

\newpage
\section{Results}\label{sec:Results}
\subsection{Fitting procedure}

The model contains 12 parameters, a large part of which is fixed either by their standard values (e.g. those of Regge trajectories, except for the Pomeron slope, whose slope exceeds the "standard" value to meet the SD data) or are set close to the previous fits \cite{PR}.

\begin{figure}[!ht]
 \includegraphics[width=0.49\linewidth,bb=10mm 24mm 195mm 185mm,clip]{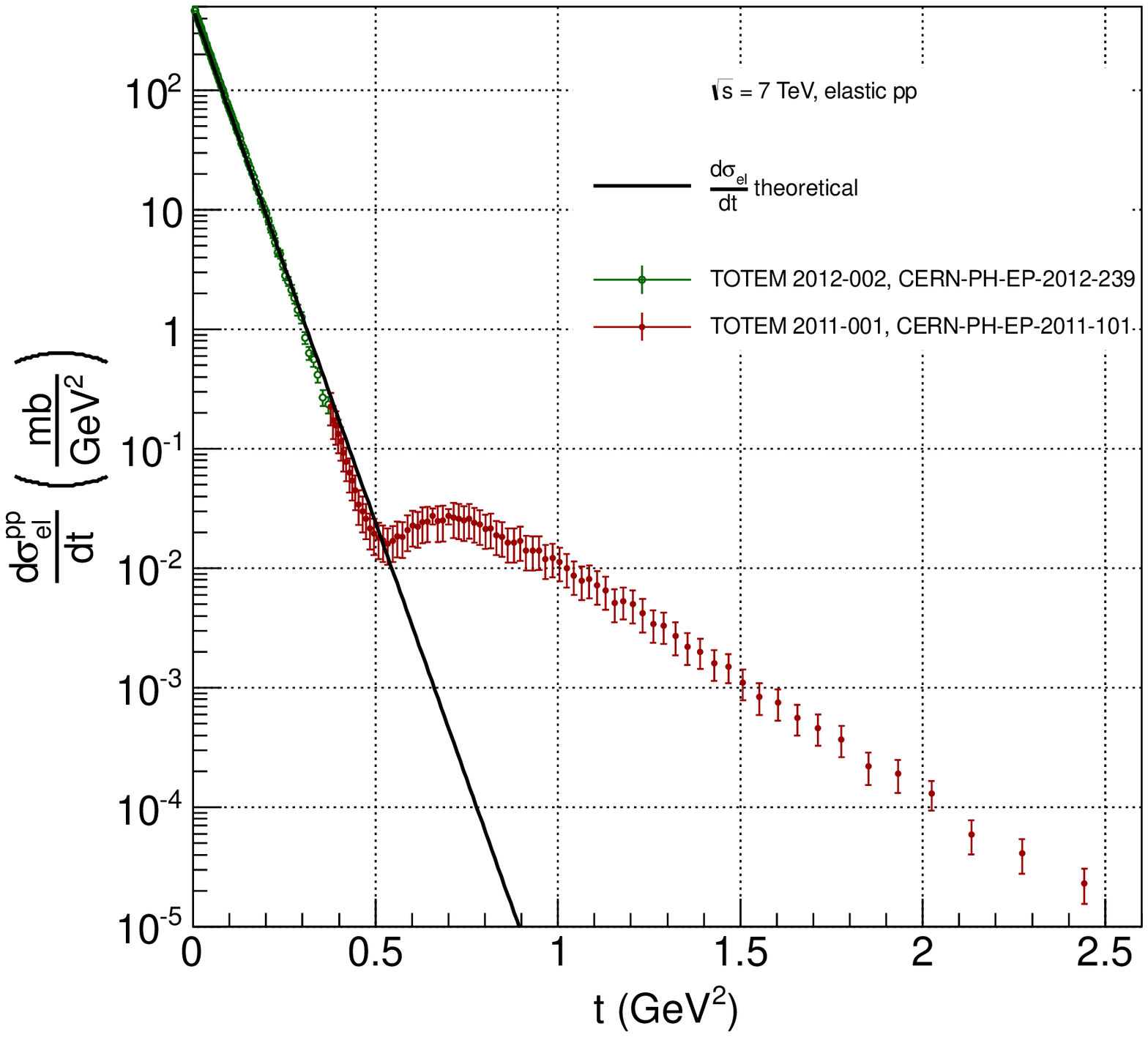}
 \includegraphics[width=0.49\linewidth,bb=10mm 24mm 195mm 185mm,clip]{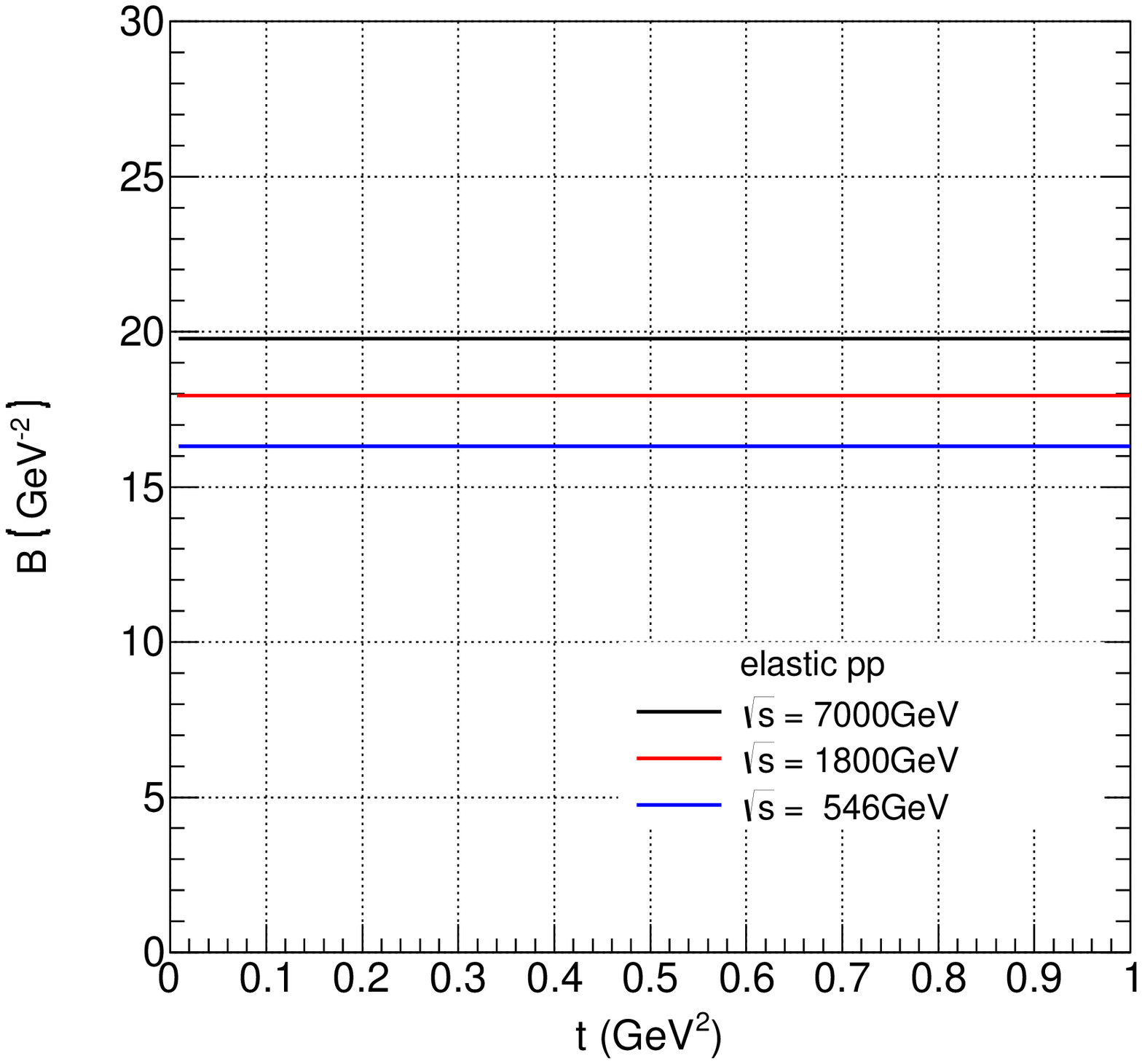}
 {(a)\hspace{0.47\linewidth}(b)}
 \caption{(a) Elastic cross section (black line) caclulated from Eq.~(\ref{elastic}) and the TOTEM data points \cite{[T1].TOTEM} at $\sqrt{s}=7$~TeV; $\chi^2/$ndf$=1.6$.\\
  (b) Slope of the elastic cross section $B=\frac{d}{dt}\left(\ln\left(\frac{d\sigma_{el}}{dt} \right)\right)$.}
 \label{fig:elastic}
\end{figure}

\begin{figure}[!ht]
 \includegraphics[width=0.49\linewidth,bb=13mm 24mm 195mm 185mm,clip]{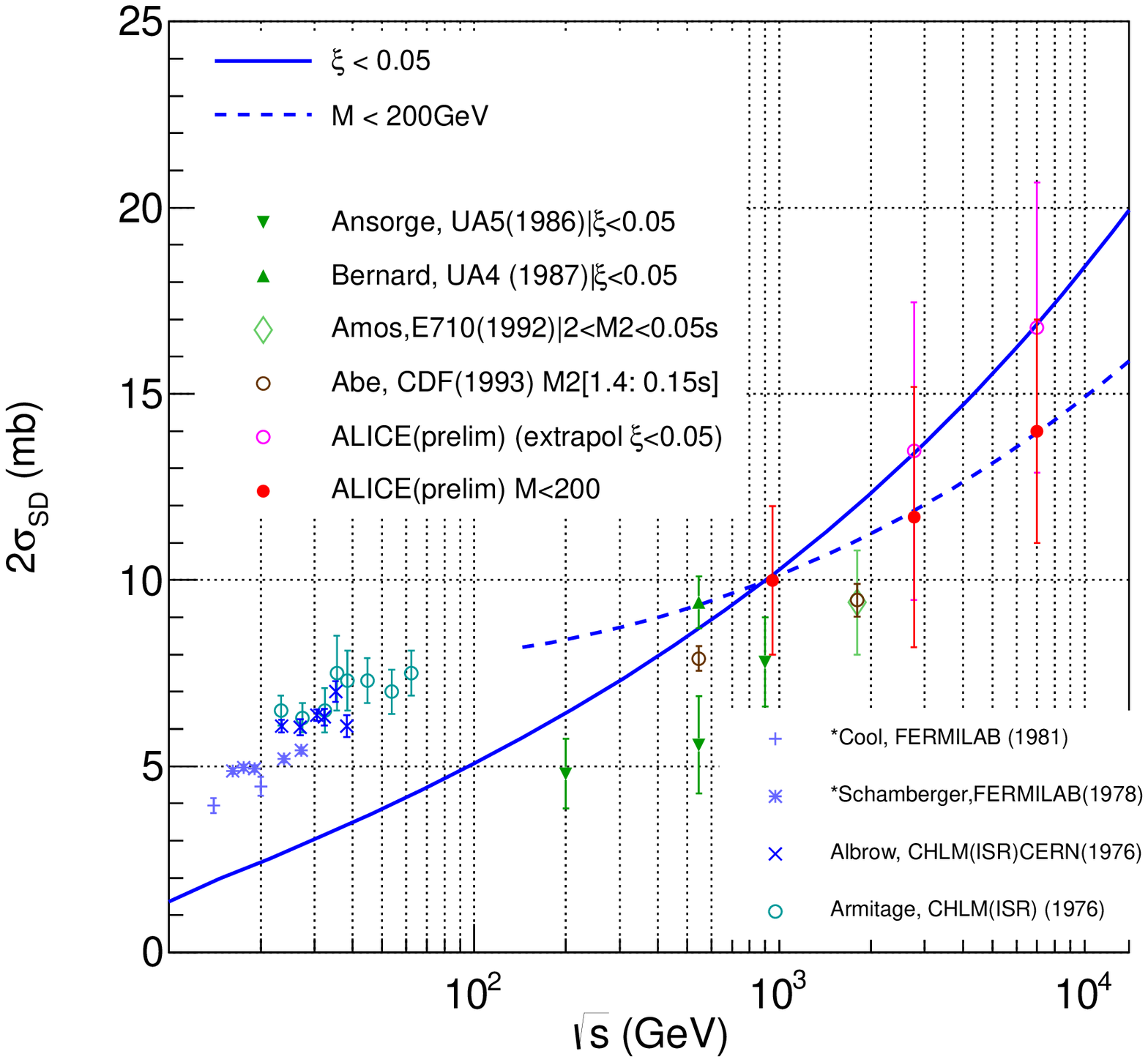}
  \includegraphics[width=0.49\linewidth,bb=13mm 24mm 195mm 185mm,clip]{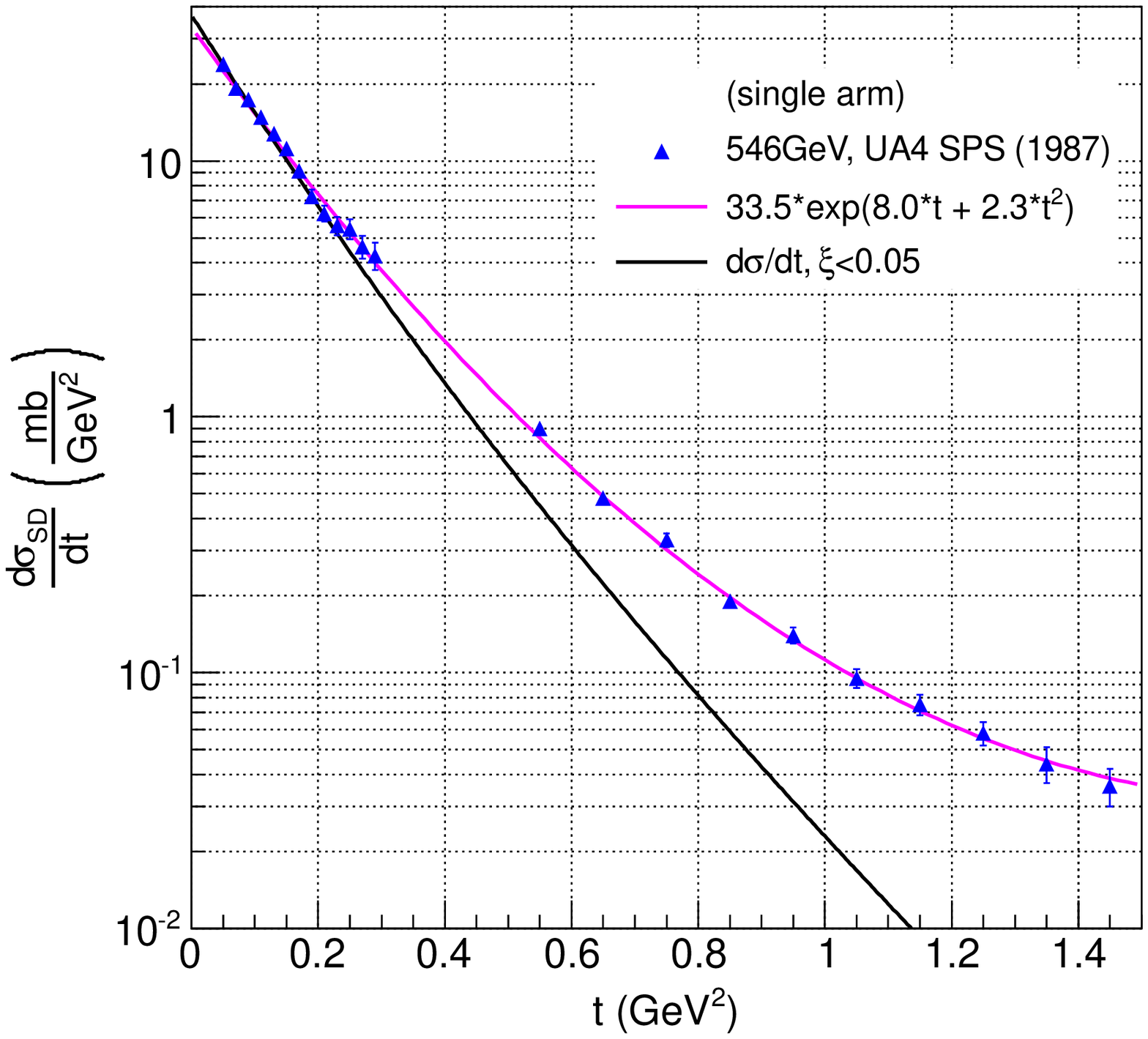}
  {(a)\hspace{0.47\linewidth}(b)}
  \caption{(a) Single diffraction dissociation cross section vs. energy $\sqrt{s}$ calculated from Eq.~(\ref{SD}). The points with red full circles correspond to (unpublished) the ALICE data ($M<200$~GeV), the point with pink open circles are those extrapolated to $\xi<0.05$ \cite{Poghosyan for ALICE}. For $\sigma|_{\xi<0.05}$: $\chi^2/$n$=1.6$, n$=6$ (only $\sqrt{s}>100$~GeV) and for $\sigma|_{M<200\mbox{~GeV}}$: $\chi^2/$n$<0.01$, n$=3$.\\
  (b) Single differential cross section $\frac{d\sigma_{SD}}{dt}$ for SD calculated from Eq.~(\ref{SD}), integrated over the region $M_x^2<0.05s$.}
 \label{fig:cs|dcsdt.SD.Data}
\end{figure}

\begin{figure}[!ht]
  \includegraphics[width=0.49\linewidth,bb=13mm 24mm 195mm 185mm,clip]{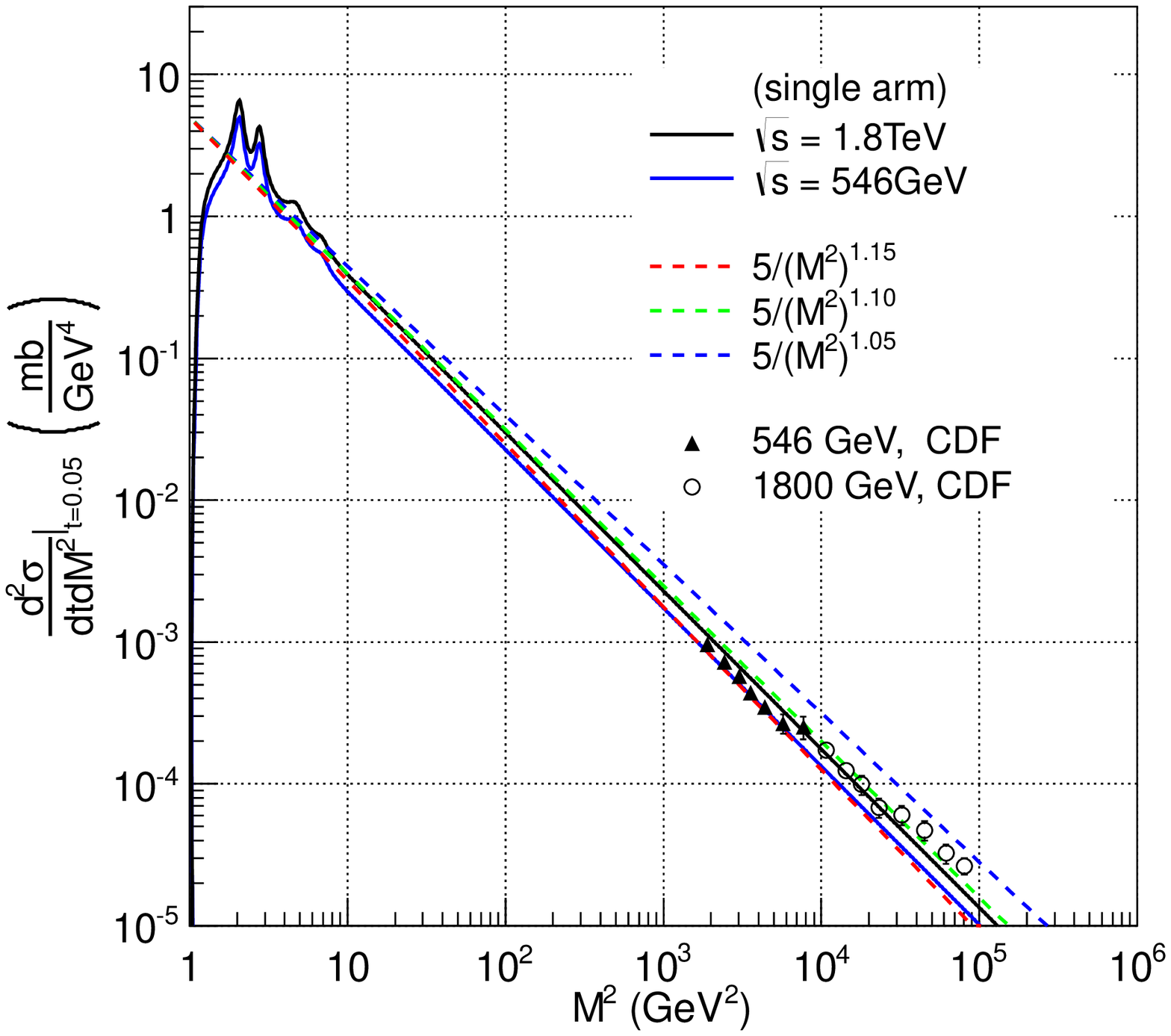}
  \includegraphics[width=0.49\linewidth,bb=13mm 24mm 195mm 185mm,clip]{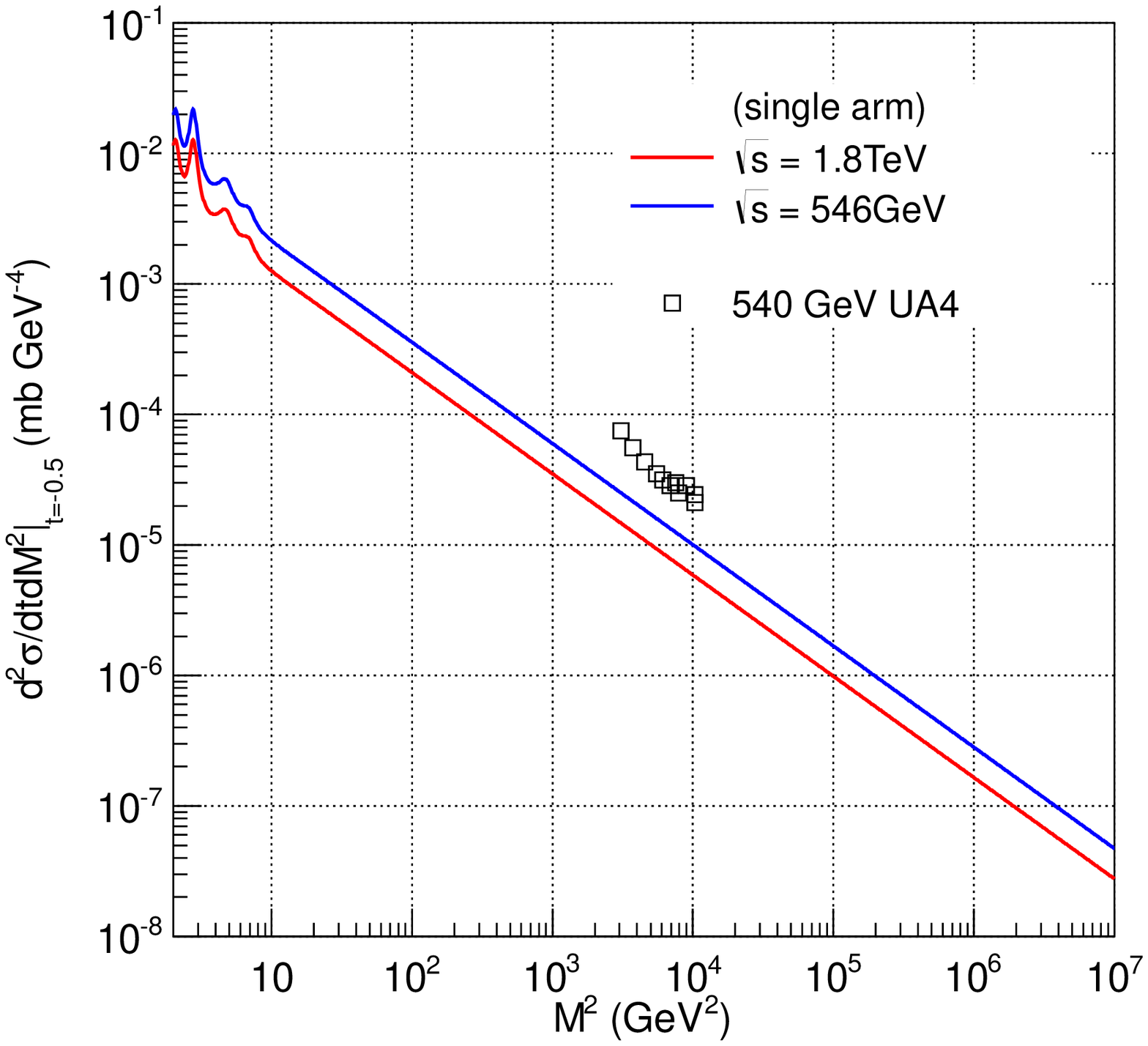}
   {(a)\hspace{0.47\linewidth}(b)}
  \caption{Double differential cross sections for SD at $t=-0.05$ (a) and $t=-0.5$ (b) calculated from Eq.~(\ref{SD}); $\chi^2/$n$=1.2$, n$=8$ for $1800$~GeV$^2$ and $\chi^2/$n$=4.6$, n$=8$ for $546$~GeV$^2$  (only Fig.~(a));}
 \label{fig:d3cs.SD.Data}
\end{figure}

\begin{figure}[!ht]
 \centering
 \includegraphics[width=0.49\linewidth,bb=13mm 24mm 195mm 185mm,clip]{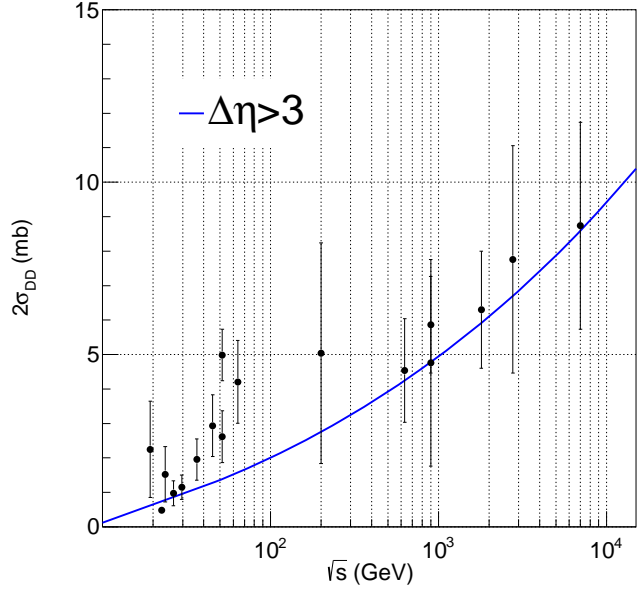}
 \caption{Double diffraction dissociation cross section vs. energy $\sqrt{s}$ calculated from Eq.~(\ref{DD}).
 Experimental data are taken from \cite{Poghosyan for ALICE}; $\chi^2/$n$=0.2$, n$=7$, for $\sqrt{s}>100$~GeV only.}
 \label{cs_s.DD}
\end{figure}

\begin{itemize}
 \item In our strategy we  first adjust the model to the "standard candles" of {\bf elastic $pp$ scattering} at high energies (starting form $500$~GeV). We considered only the data corresponding to the first cone, described by a linear exponential, implying also a linear Pomeron trajectory.

 The elastic data and Regge theory fix the parameters  $s_0, \alpha(0)$, $\alpha'$, $A_{el}$, $b_{el}$. The relevant curves, the data and values of the fitted parameters are shown in Fig.~\ref{fig:elastic}(a), in Tables \ref{tab:cs.el} and \ref{tab:Bslope.el}.

The  data at larger $|t|$, with the dip-bump structure and subsequent flattening of the cross section, both in elastic scattering and in SD may indicate the onset of new physics and the transition to hard scattering, implying a non-exponential residue and/or a non-linear Pomeron trajectory, see Refs. \cite{JLL}, that goes beyond the present study.

{\it{\bf NB}: The parameter $s_0$ is strongly correlated with the slope parameters $b_{el}$, $b_{res}$ and $b_bg$.}

\item {\bf Single diffraction dissociation (SD)} is an important pillar in our fitting procedure. The following parameters were fitted to SD data: $A_{res}$, $C_{bg}$, $R$, $b_{res}$, $b_{bg}$, $\varsigma$, $\eta$. As input data we used: a)double differential cross sections $\frac{d^2\sigma_{SD}}{dtdM^2}$ versus $M_x^2$ at $|t|=0.05$~GeV$^2$ (see Fig.~\ref{fig:d3cs.SD.Data}(a)) and b) at at $|t|=0.5$~GeV$^2$ (see Fig.~\ref{fig:d3cs.SD.Data}(b)); c)single differential cross sections $\frac{d\sigma_{SD}}{dt}$ vs. $t$ (see Fig.~\ref{fig:cs|dcsdt.SD.Data}(b)); d) fully integrated cross sections versus energy $\sqrt{s}$ (see Fig.~\ref{fig:cs|dcsdt.SD.Data}(a)).

In particular due to the difference between the kinematical regions chosen by the different experimental groups, there are some discrepancies int the data. 

 At low $t$ (below $0.5$~GeV$^2$), the $t-$ dependence of SD cross section are well described by an exponential fit, see Figs. \ref{fig:cs|dcsdt.SD.Data}(b) and \ref{fig:d3cs.SD.Data}(b)), but beyond this region the cross sections start flattening due to transition effects towards hard physics.

 Low-energy data $\sqrt{s}<100$~GeV (see Figs. \ref{fig:cs|dcsdt.SD.Data}(a), \ref{cs_s.DD}) require the inclusion of non-leading Reggeons, so they are outside our single Pomeron exchange in the $t$ channel.

 \item {\bf Double diffraction dissociation (DD) cross sections} follow, up to some fine-tuning of the parameters, from our fits to SD and factorization relation.


 Integration in $M_1^2$ and $M_2^2$ comprises the range $\Delta\eta>3$, where $\Delta\eta=\ln\left(\frac{ss_0}{M_1^2M_2^2}\right)$.
\end{itemize}




The $\chi^2$ values are quoted at relevant figures.  
The values of the fitted parameters are presented in Table \ref{tab:ParSet} and
our prediction  are summarize in Table \ref{tab:cs.predict}.
~~~~~~~~~~~~~~~~~~~~~~~~~~~~~~~~~~~~~~~~~~~~~~~~~~~~~~~~~

\begin{table}[!ht]
 \begin{tabular}{ccc}
    \begin{tabular}{|c|c|}
      \hline
                       &Data [GeV$^{-2}$]                       \\ \hline
       $B_{el}(7$~TeV$)$  &$19.9 \pm0.3$ \cite{[T1].TOTEM}         \\ \hline
       $B_{el}(1.8$~TeV$)$&$17.0 \pm0.5$ \cite{[tmp3]Amos.E710.[3]}\\
                       &$17.9 \pm2.5$ \cite{[tmp3]Amos.E710.[5]}\\
                       &$16.99$ \cite{[tmp4]Abe.CDF}            \\ \hline
       $B_{el}(546$~GeV$)$&$15.35$ \cite{[tmp4]Abe.CDF}            \\
                       &$15.0$ \cite{[tmp6].Ansorge.UA5}        \\ \hline
   \end{tabular}&\qquad\qquad&
   \begin{tabular}{|c|c|c|}
      \hline
                            &Data [$mb$]                             & Calcuation [$mb$]\\ \hline
       $\sigma_{el}(7$~TeV$)$  &$25.4\pm1.1$ \cite{[T1].TOTEM}          &   24.5   \\ \hline
       $\sigma_{el}(1.8$~TeV$)$&$16.6\pm1.6$ \cite{[tmp3]Amos.E710.[3]} &   17.99  \\ \hline
       $\sigma_{el}(546$~GeV$)$&$13.6$      \cite{Bozzo1}               &   13.8   \\ \hline
   \end{tabular}\\
   \parbox[t]{6.cm}{\caption{\label{tab:Bslope.el} Forward slope of elastic $pp$ scattering, see Fig.~\ref{fig:elastic}(b)}}& &
   \parbox[t]{8.cm}{\caption{\label{tab:cs.el} $pp$ elastic cross section, Eq.~(\ref{elastic}), calculated with the parameters quoted in Tab. \ref{tab:ParSet}.}}\\
 \end{tabular}
\end{table}


\newpage
\begin{table}[!ht]
\centering
  \begin{tabular}{ccc}
     \begin{tabular}{|c||c|}
        \hline
        $A_{el}[mb]$        & $33.579$\\
        $b_{el}[$GeV$^{-2}]$  &$1.937$\\ \hline

    $A_{res}[mb\cdot$GeV$^4]$&$2.21$\\
        $C_{bg}[mb]$        &$2.07$\\
        $R$                 &$0.45$\\
        $b_{res}[$GeV$^{-2}]$ &$-0.507$\\
        $b_{bg}[$GeV$^{-2}]$  &$-1.013$\\ \hline

     \end{tabular}&\qquad\qquad&
      \begin{tabular}{|c||c|}
        \hline
        $s_0$   &$1$\\ \hline
        $\varsigma$ &$0.8$\\
        $\eta$      &$1$\\ \hline
        \hline
        \multicolumn{2}{|c|}{$\alpha(t)=\alpha(0)+\alpha't$}\\ \hline
        $\alpha(0)$           & $1.075$\\
        $\alpha'$[GeV$^{-2}$] & $0.34$\\\hline
     \end{tabular}%
  \end{tabular}%
 \caption{\label{tab:ParSet} Fitted parameters, see Eqs.~(\ref{elastic}), (\ref{SD}), (\ref{DD}).}
\end{table}

\begin{table}[!ht]
  \begin{tabular}{c c c}
    \begin{tabular}{|c||c|}
      \hline
      $\sigma_{SD}|_{\xi<0.05}$ & $16.7$\\
      $\sigma_{SD}|_{M<4\mbox{~GeV}}$&  $4.9$\\
      $\sigma_{SD}|_{M>4\mbox{~GeV}}$&  $11.8$\\ \hline \hline
      $\sigma_{DD}|_{\Delta\eta>3}$      & $8.6$\\
      $\sigma_{DD}|_{M^2_1M^2_2<16\mbox{~GeV}^2}$& $0.4$\\
      $\sigma_{DD}|_{M^2_1M^2_2>16\mbox{~GeV}^2}$& $8.2$\\ \hline
    \end{tabular}&\qquad&

    \begin{tabular}{|c||c|c||c|}
       \hline
       $\sqrt{s} [$~TeV$]$& $\sigma_{SD}|_{M^2<200GeV}$& $\sigma_{SD}|_{\xi<0.05}$& $\sigma_{DD}|_{\Delta\eta>3}$\\  \hline
       7 &13.7& 16.7& 8.6 \\
       10&14.6& 18.2& 9.4 \\
       14&15.5& 19.7&10.2 \\ \hline
    \end{tabular}\\
    
   \parbox[t]{6.cm}{\caption{\label{tab:cs.SD.DD.res} Cross sections [mb] calculated at 7~TeV.}}& &
   \parbox[t]{9.cm}{\caption{\label{tab:cs.predict}Predictions (in [mb]) for SD and DD at 7, 10 and 14~TeV.}}\\
    
 \end{tabular}
\end{table}


\subsection{pp elastic}
\newpage
\subsection{Single diffraction dissociation}

\newpage
\begin{figure}[!ht]
 \includegraphics[width=0.49\linewidth,bb=13mm 24mm 195mm 185mm,clip]{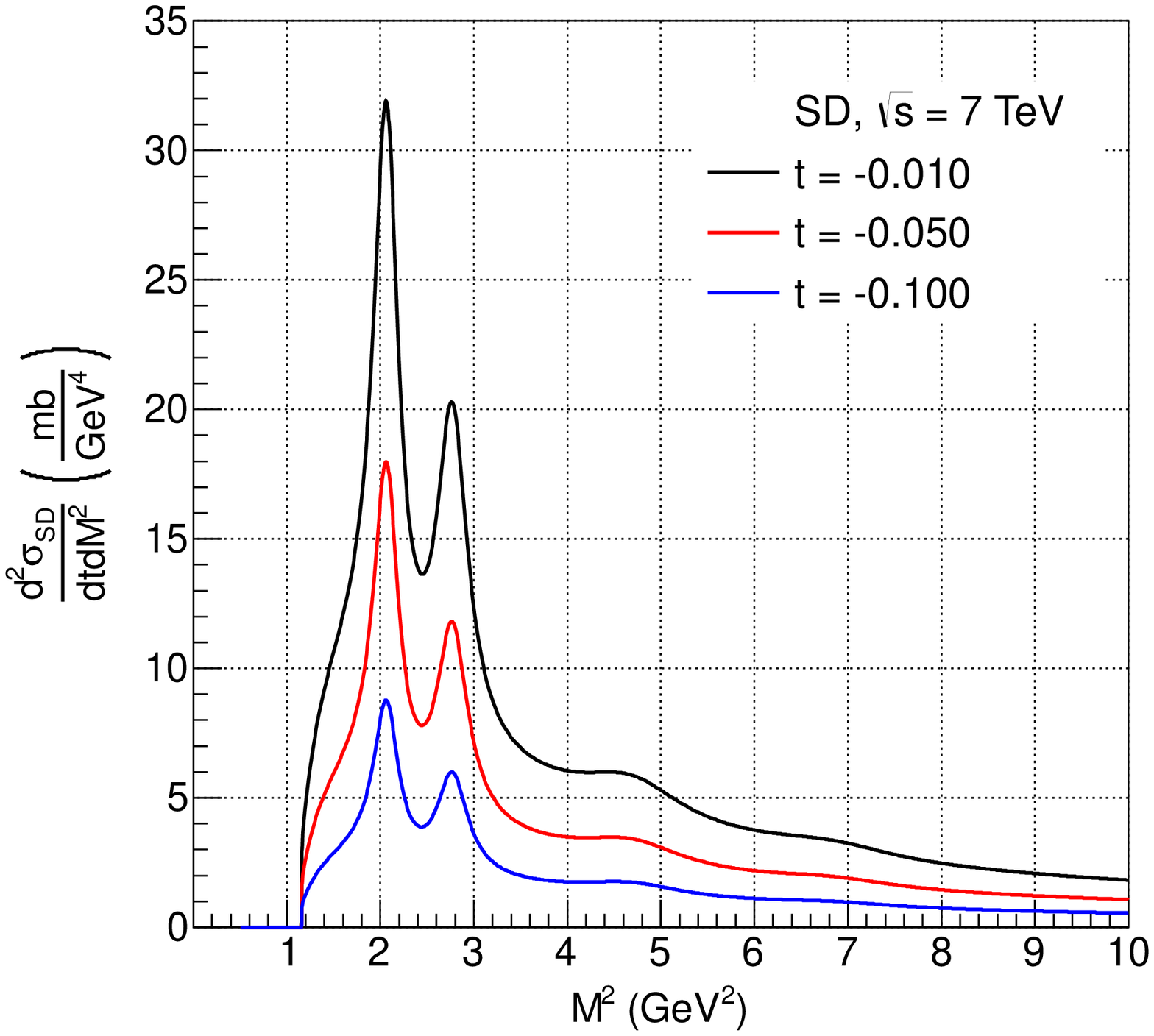}
 \includegraphics[width=0.49\linewidth,bb=13mm 24mm 195mm 185mm,clip]{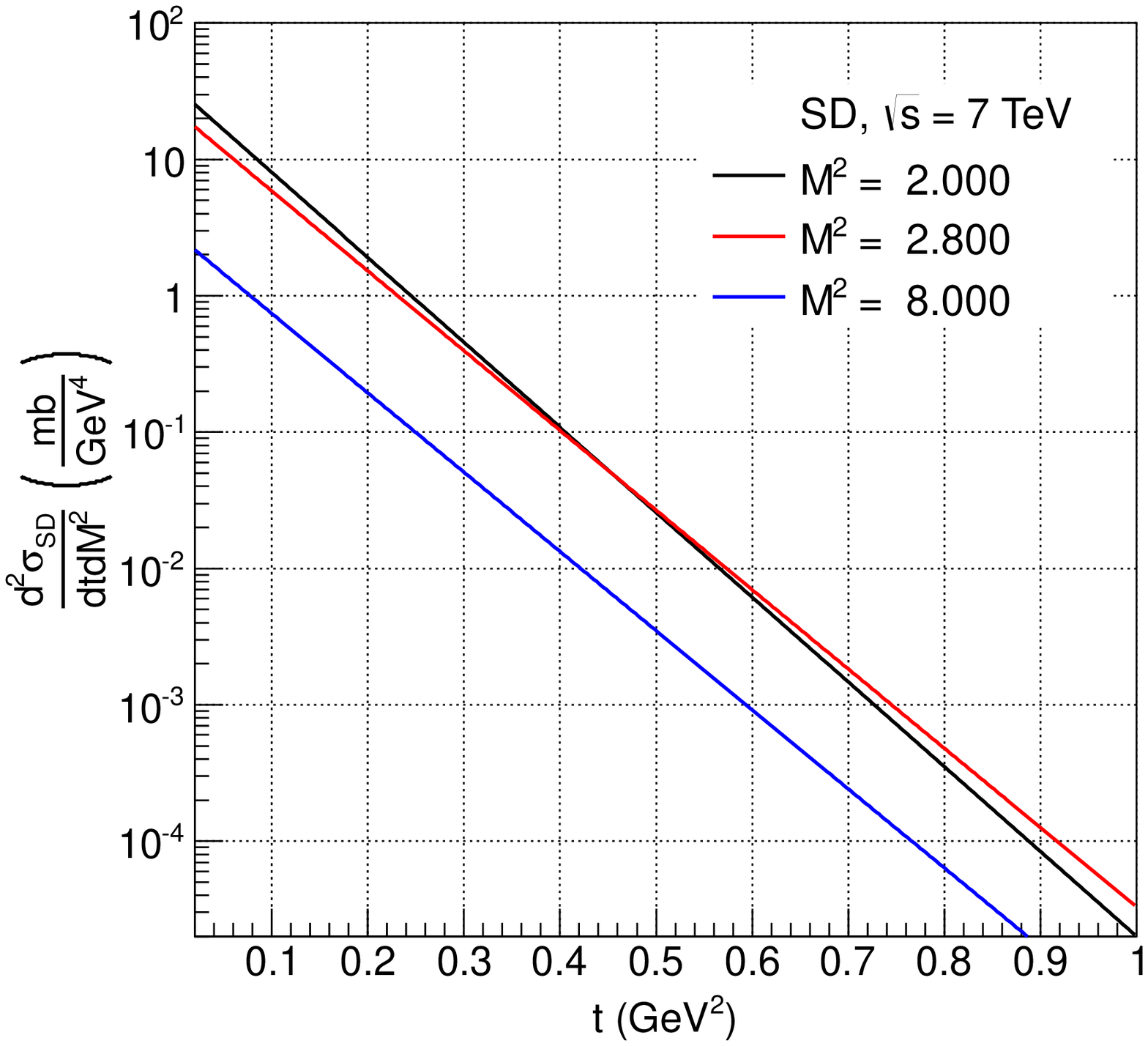}
  {(a)\hspace{0.47\linewidth}(b)}
 \caption{(a) Double differential SD cross sections as functions of $M^2$ for different $t$ values, see Eq.~(\ref{SD}).\\
   (b) Double differential SD cross sections as functions of $t$ for different $M^2$ values, see Eq.~(\ref{SD}).}
 \label{d2cs_m2.in.t}
\end{figure}

\begin{figure}[!ht]
 \includegraphics[width=0.49\linewidth,bb=13mm 24mm 195mm 185mm,clip]{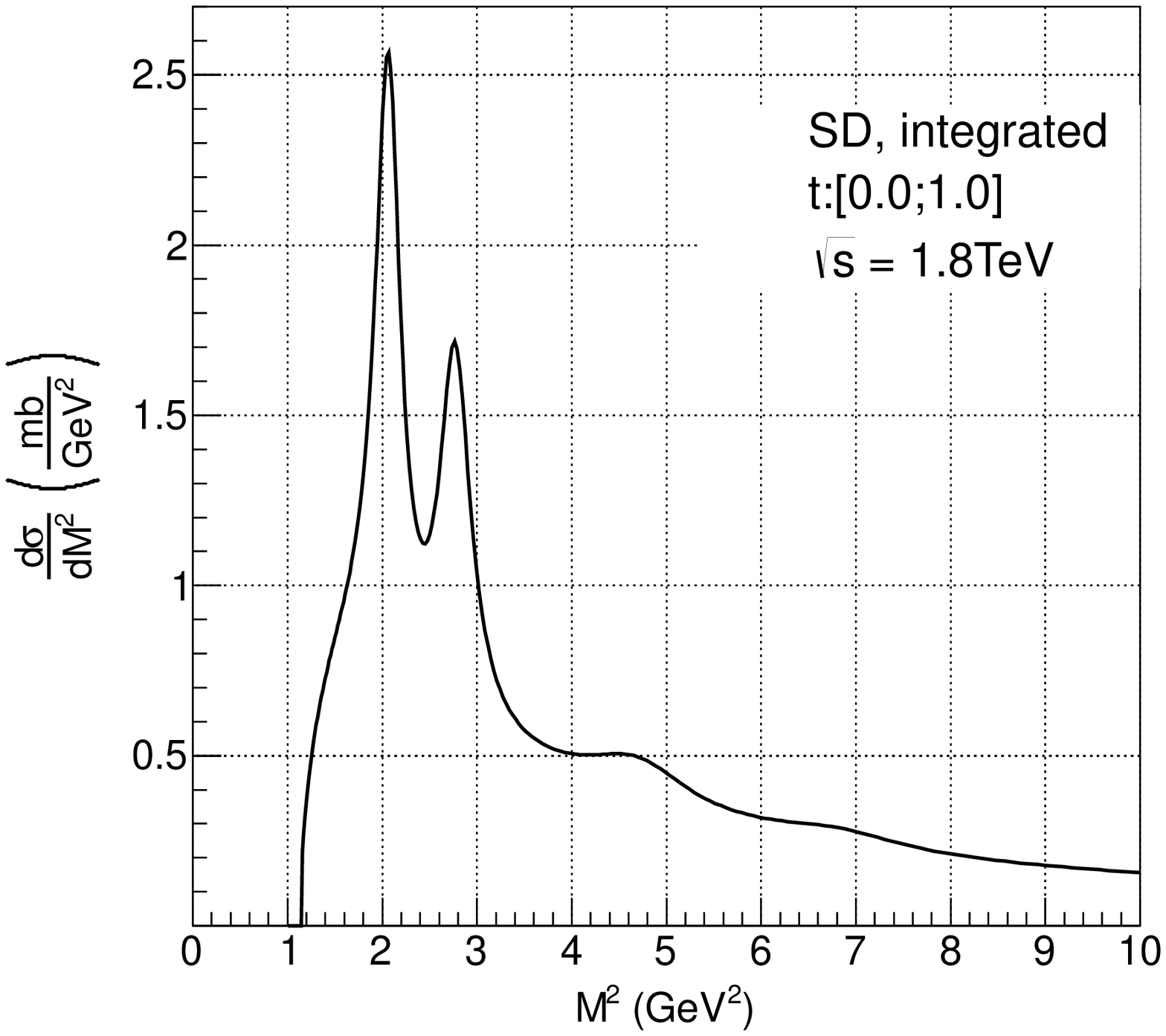}
 \includegraphics[width=0.49\linewidth,bb=13mm 24mm 196mm 185mm,clip]{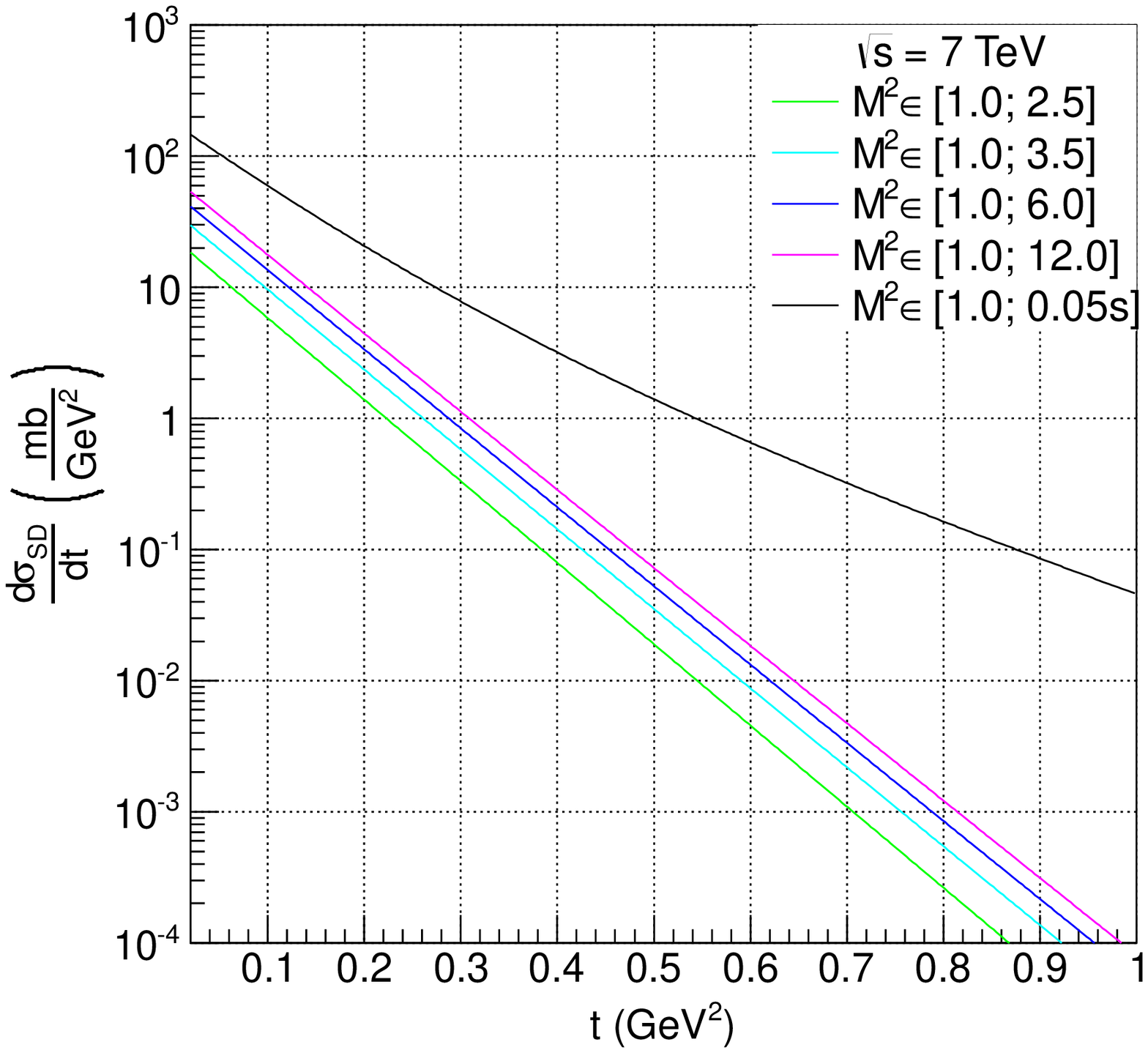}
  {(a)\hspace{0.47\linewidth}(b)}
 \caption{(a) Single differential SD cross section as a function of $M^2$ integrated over $t\in[0.0:1.0]$, see Eq.~(\ref{SD}).\\
  (b) Single differential SD cross section as a function of $t$, integrated over the $M_x^2$ from the threshold up to some fixed values, see Eq.~(\ref{SD}).}
 \label{int.dcsdM2}
\end{figure}

\newpage
\begin{figure}[!ht]
 \includegraphics[width=0.49\linewidth,bb=13mm 24mm 195mm 185mm,clip]{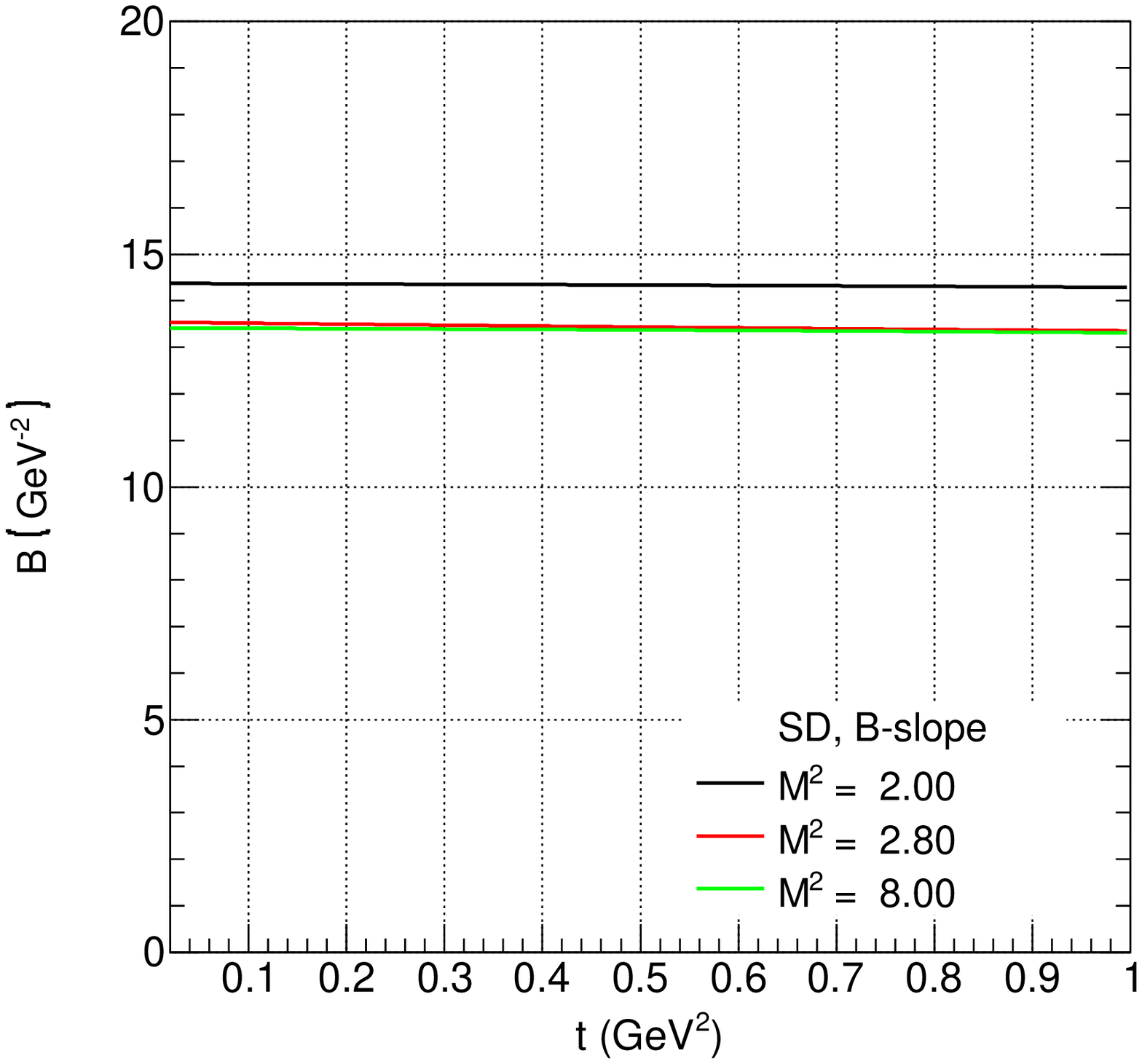}
 \includegraphics[width=0.49\linewidth,bb=13mm 24mm 195mm 185mm,clip]{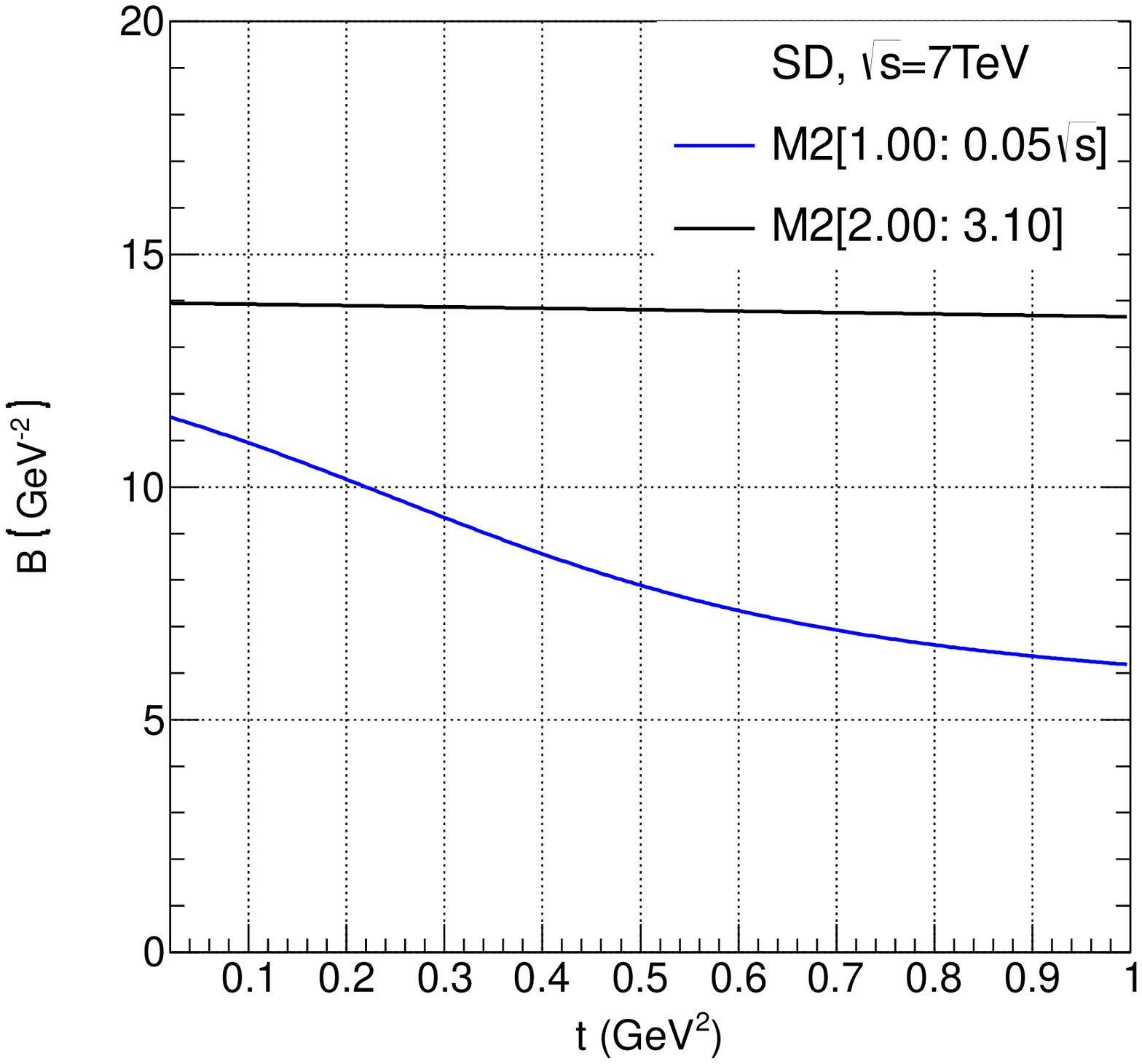}
  {(a)\hspace{0.47\linewidth}(b)}
 \caption{(a) Slopes of differential SD cross section at different $M^2$ values,
 $B=\frac{d}{dt}\ln\left.\frac{d^2\sigma_{SD}}{dtdM_x^2}\right|_{M^2_x=\{M^2_1, M^2_2, M^2_3\}}$.\\
 (b) Slopes of differential SD cross section integrated in $M_x^2$ over the region of the first resonance, then over the whole diffraction region. $B=\frac{d}{dt}\ln\frac{d\sigma_{SD}}{dt}$, where $\frac{d\sigma_{SD}}{dt}=\int_{M_1^2}^{M_2^2}\frac{d^2\sigma_{SD}}{dtdM_x^2}dM_x^2$.\\ Double differential SD cross section $\frac{d^2\sigma_{SD}}{dtdM_x^2}$ is calculated from Eq.~(\ref{SD}) at $\sqrt{s}=7$~TeV.}
 \label{B_Res}
\end{figure}

\begin{figure}[!ht]
 \includegraphics[width=0.49\linewidth,bb=13mm 24mm 195mm 185mm,clip]{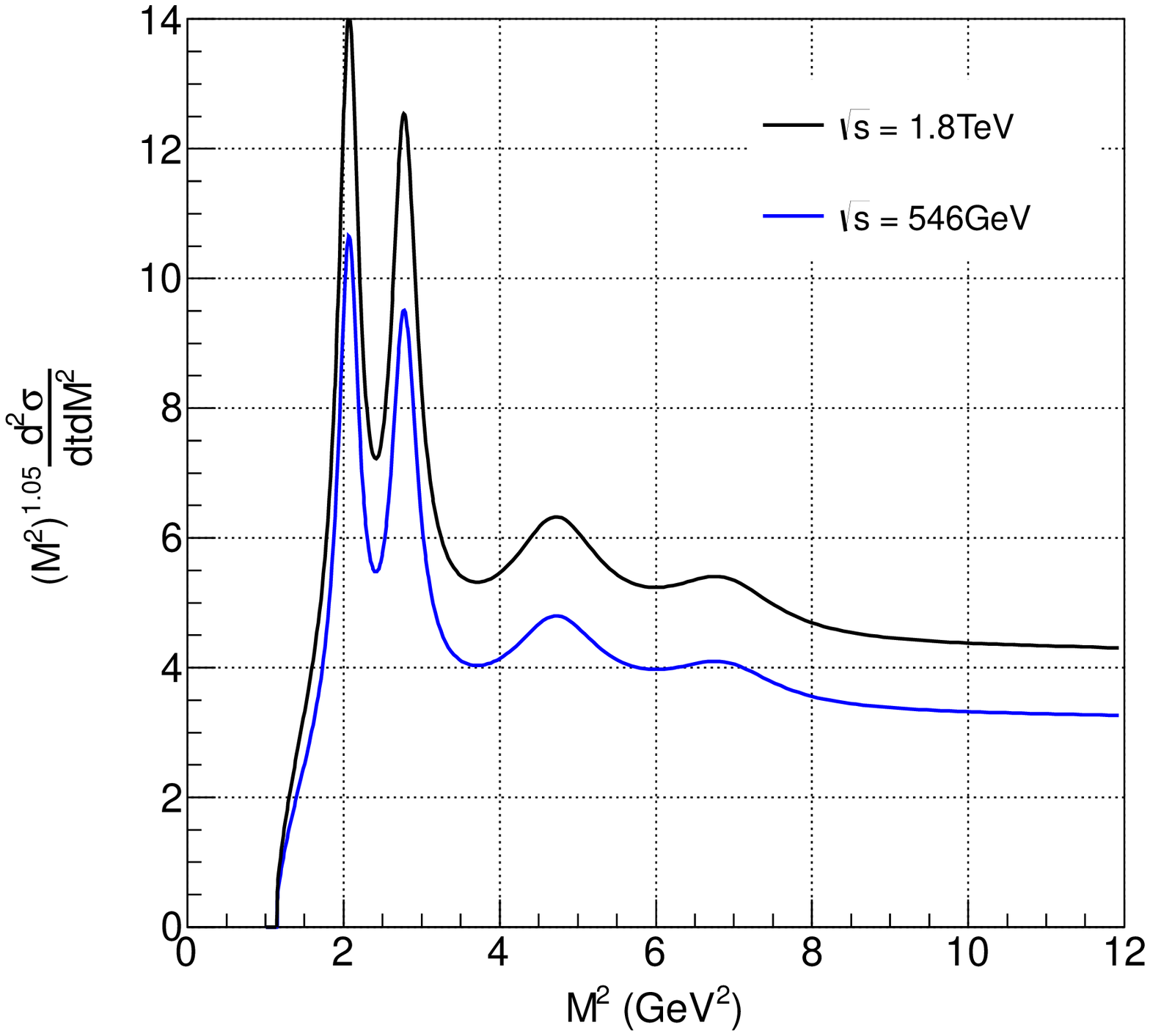}
 \includegraphics[width=0.49\linewidth,bb=13mm 24mm 196mm 185mm,clip]{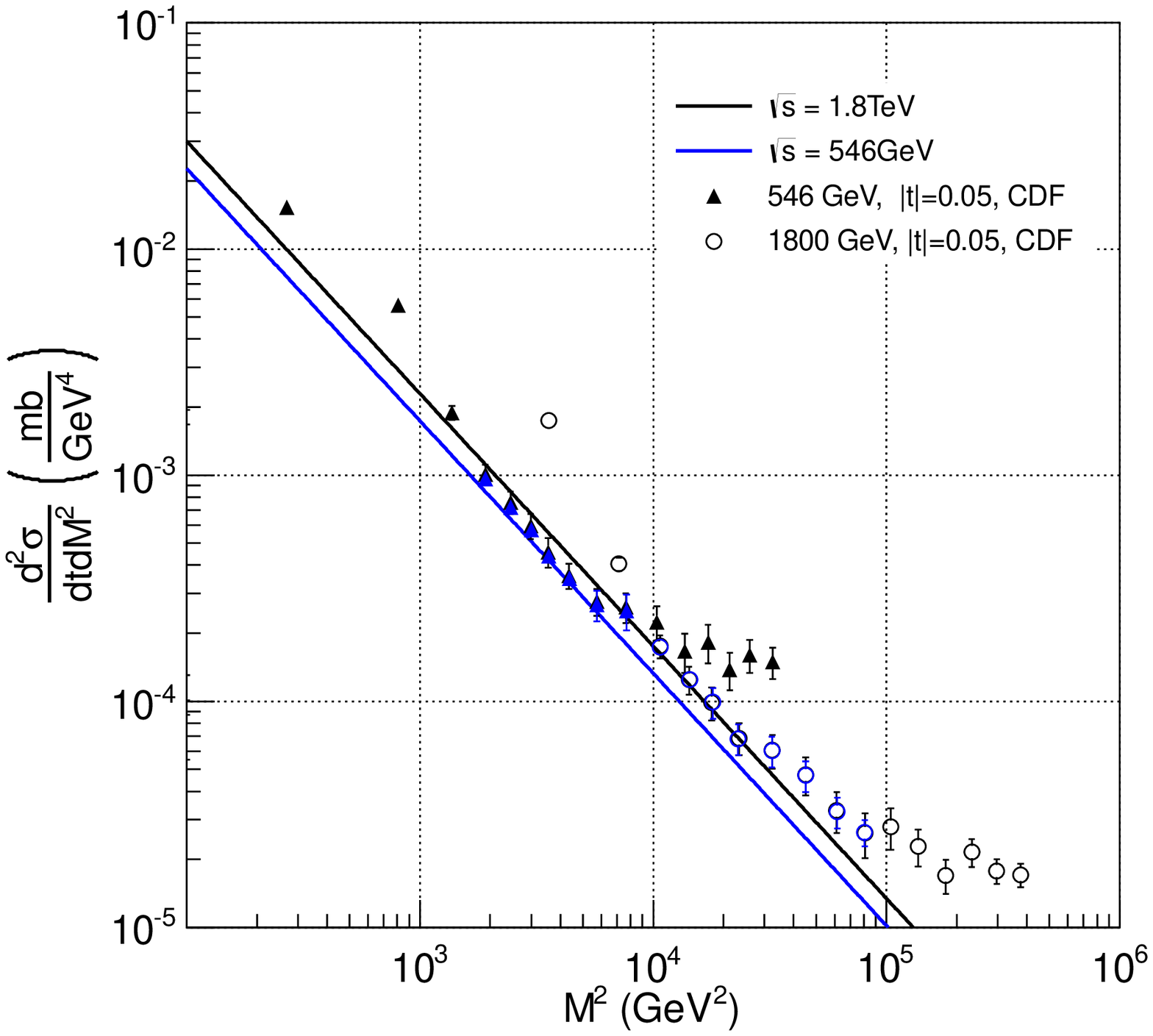}
   {(a)\hspace{0.47\linewidth}(b)}
 \caption{(a) Normalized cross sections to showing the the prominent role of the resonances over the background (cf. with the finite mass sum rules, Fig.~26 in  Ref.~\cite{Goulianos}).\\
 (b) Double differential cross section $\frac{d\sigma_{SD}}{dtdM_x^2}$ for SD at $546$ and $1800$~GeV at $|t|=0.05$.The blue points correspond to diffracion reggion $0.01<\xi<0.05$. The flattening of the cross section above $\xi>0.05$ may indicate the contribution of the Pomeron exchange in the $Pp$ total cross section. However, this region may not be diffractive any more.}
 \label{add.SD}
\end{figure}


\newpage
\subsection{Double diffraction dissociation}

\begin{figure}[!ht]
 \includegraphics[width=1\linewidth,bb=0 0 1400 900,clip]{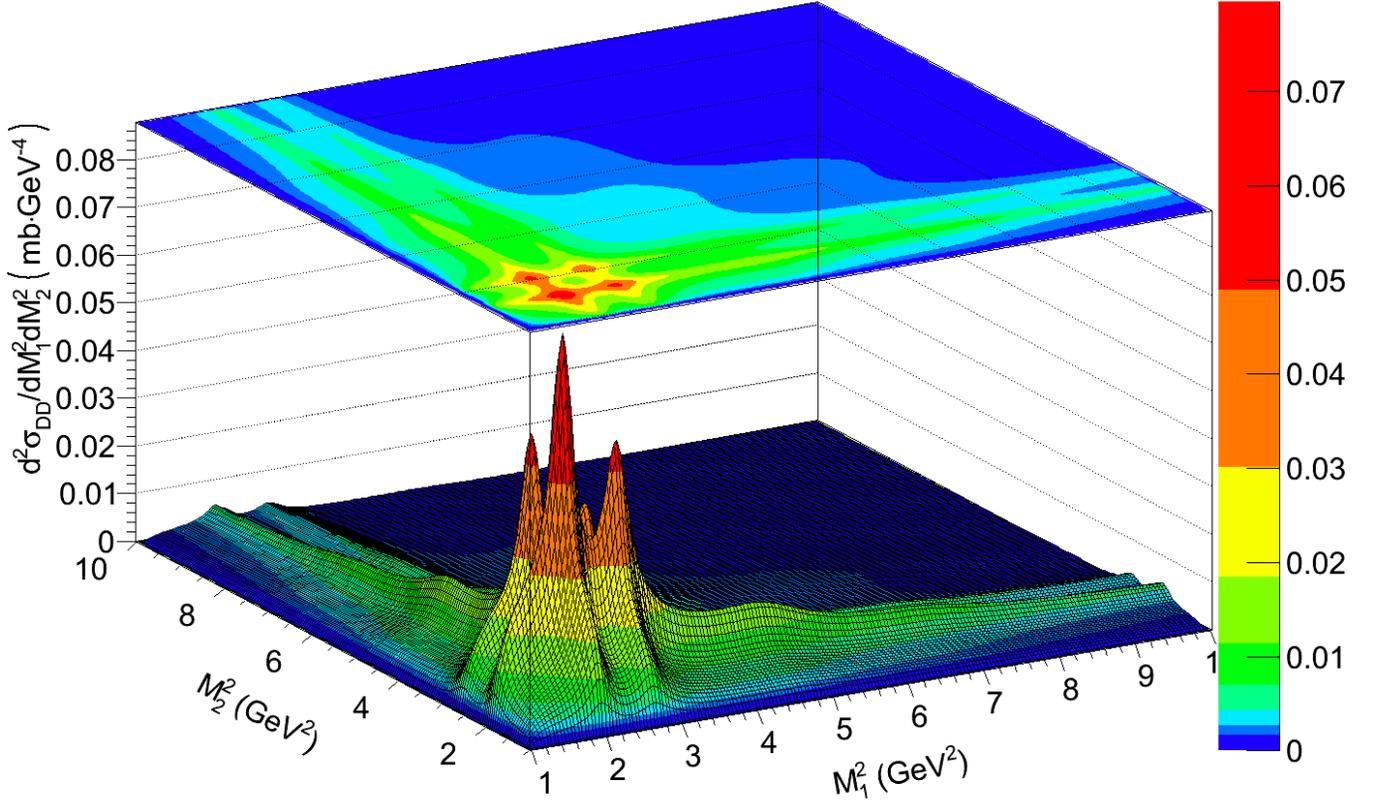}
\caption{Double differential DD cross section as a function of $M_1^2$ and $M_2^2$ integrated over $t$; see Eq.~(\ref{DD}).}\label{int.d2cs|dcs.DD}
\end{figure}

\newpage
\begin{figure}[!ht]
 \includegraphics[width=0.49\linewidth,bb=13mm 24mm 195mm 185mm,clip]{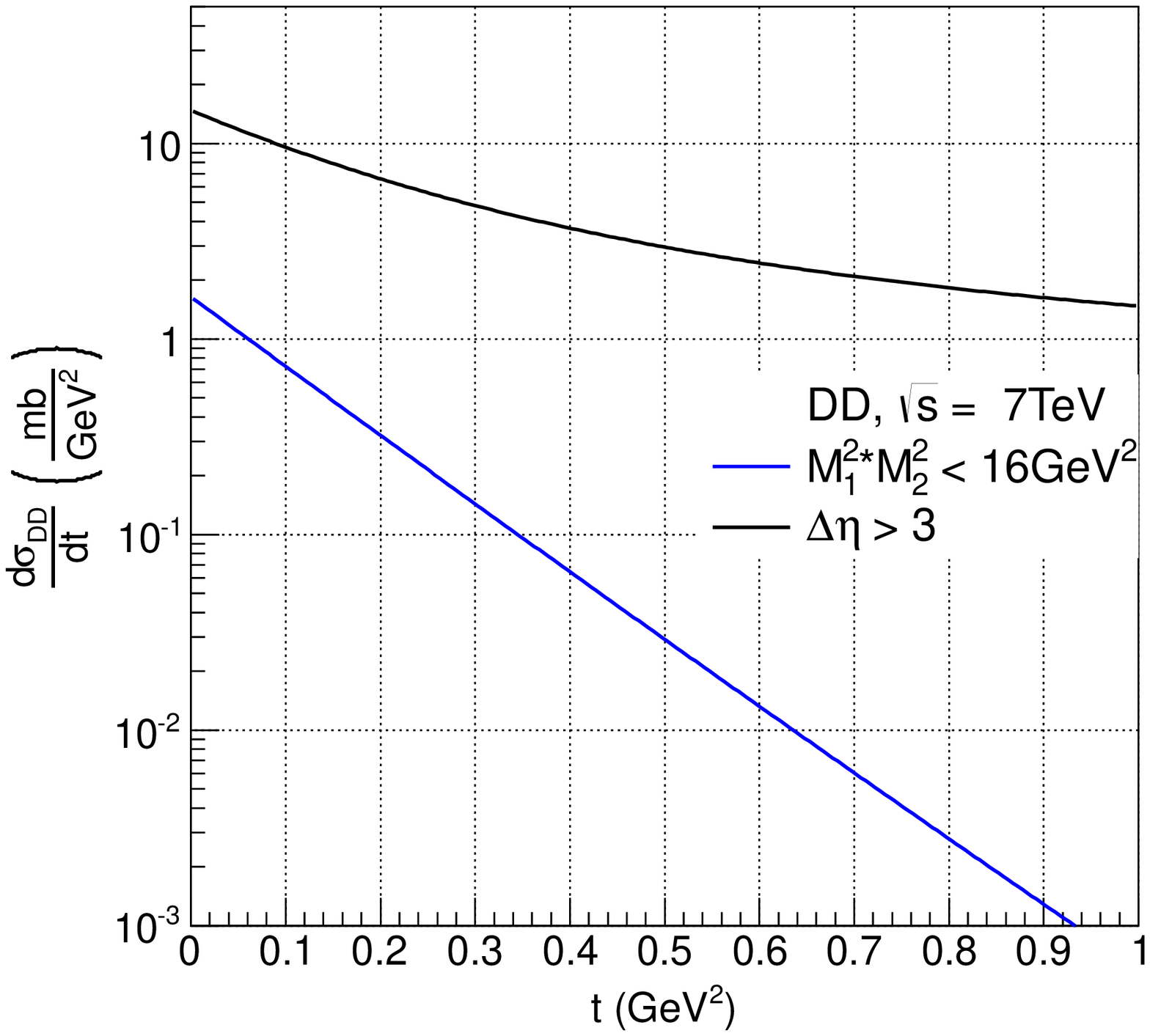}
 \includegraphics[width=0.49\linewidth,bb=13mm 24mm 195mm 185mm,clip]{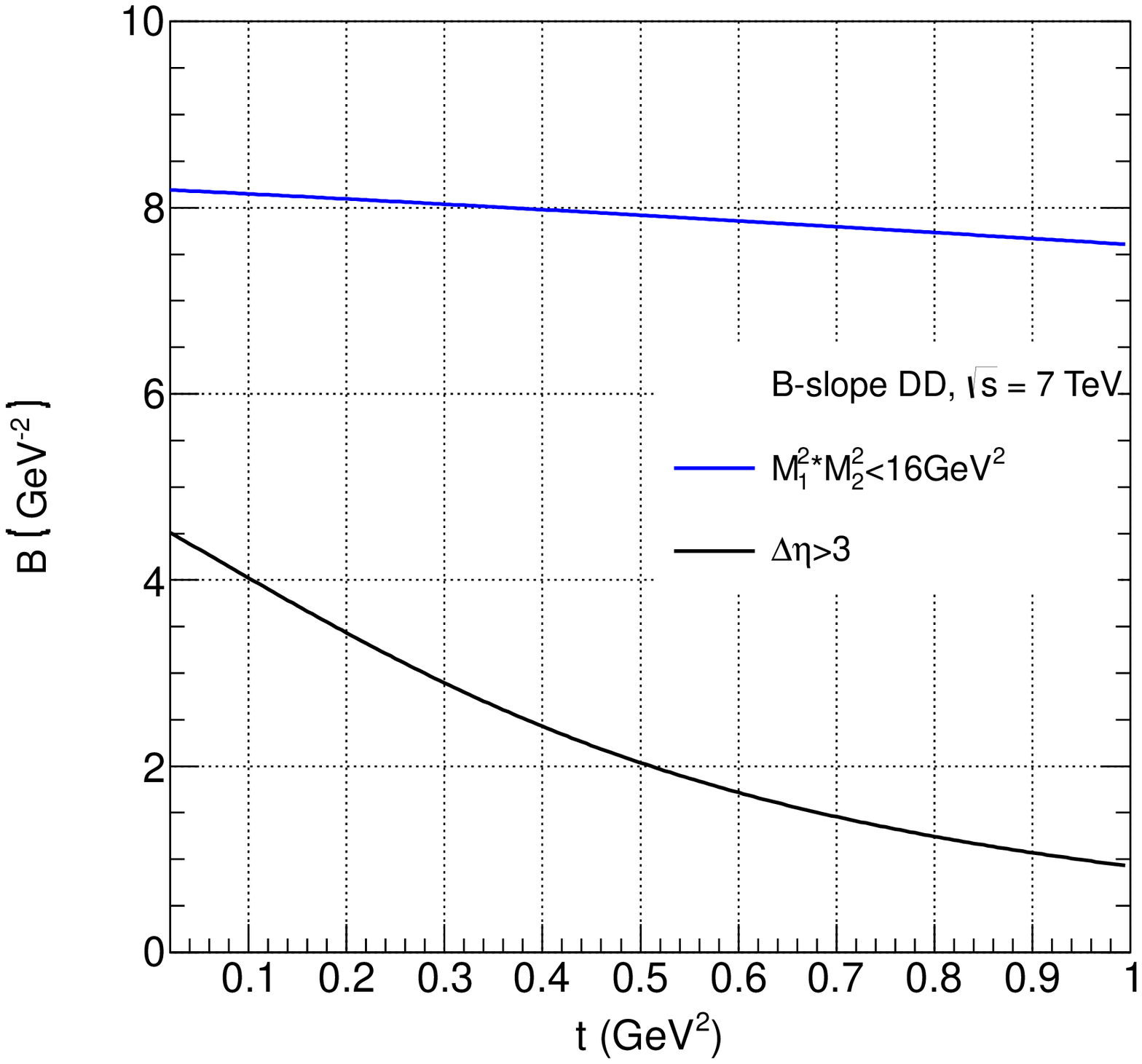}
  {(a)\hspace{0.47\linewidth}(b)}
 \caption{(a) Single differential cross section $\frac{d\sigma_{DD}}{dt}$ as a function of $t$ integrated over the region of resonances $\left(M_1^2M_2^2<16 \mbox{~GeV}^2\right)$ and over the whole region of diffraction ($\Delta\eta>3$), see Eq.~(\ref{DD}).
 (b) The slope   $B=\frac{d}{dt}\ln\left(\frac{d\sigma_{DD}}{dt}\right)$ is calculated from Eq.~(\ref{DD}).}
 \label{dcsdt|B.DD}
\end{figure}

\begin{figure}[!ht]
   \includegraphics[width=0.49\linewidth,bb=0 0 1400 900,clip]{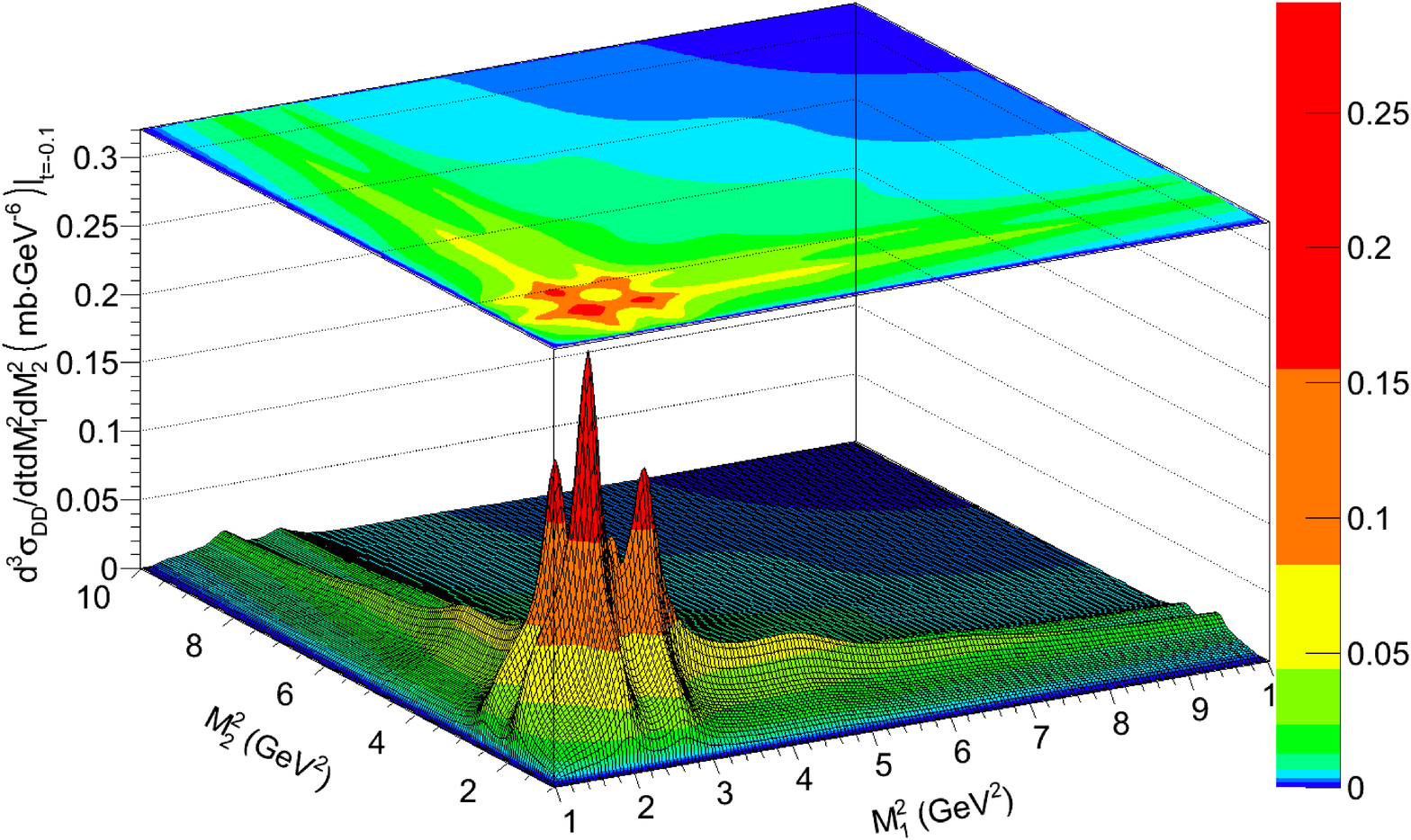}
   \includegraphics[width=0.49\linewidth,bb=0 0 1400 900,clip]{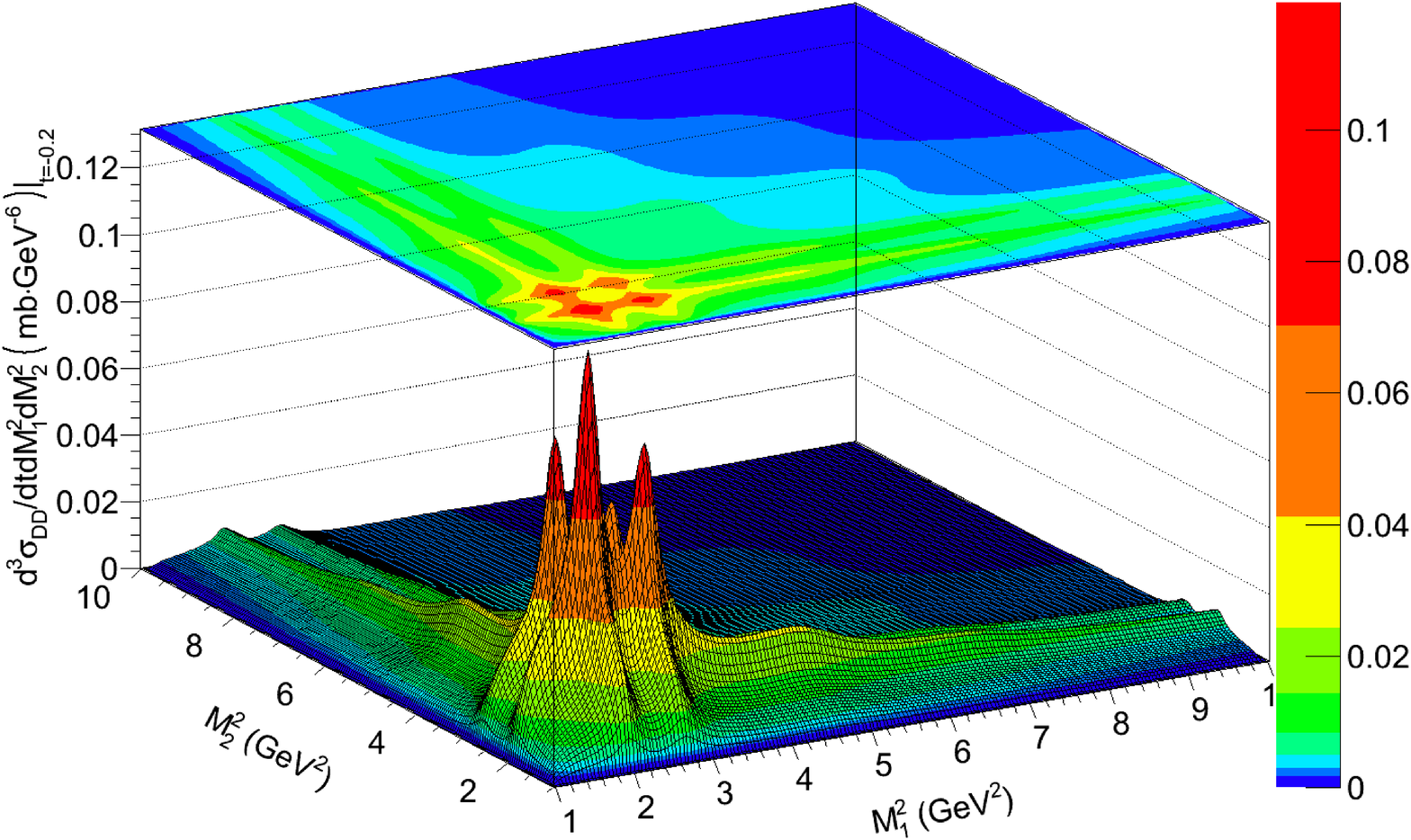}
   \includegraphics[width=0.49\linewidth,bb=0 0 1400 900,clip]{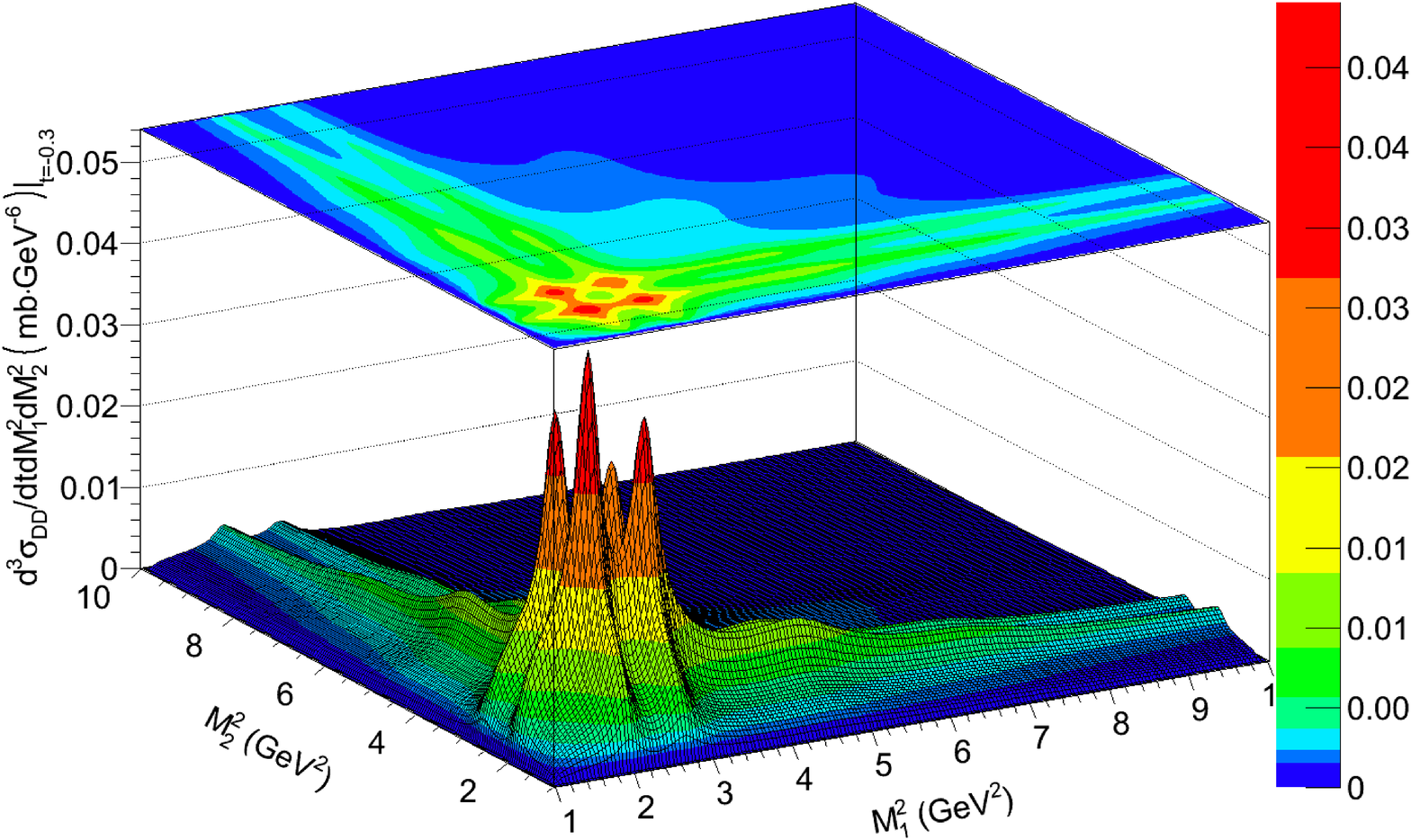}
 \caption{Triple differential DD cross sections $\frac{d\sigma_{DD}}{dtdM_1^2dM_2^2}$ as a functions of $M_1^2$ and $M_2^2$ for different $t$ values: $t \in \{-0.1; -0.2; -0.3\}$ calculated from Eq.~(\ref{DD}).}
 \label{d3cs/dtdM1dM2.DD}
\end{figure}
\newpage

\newpage
\section{Appendix: The kinematical factor $K(x_B,M_2^x)$ and turn-down of the cross sections as $t\rightarrow 0$}\label{sec:Appendix}
By guge invariance the DIS SF, and consequently our $Pp$ transicion amplitude should vanish as $Q^2$ when $Q^2\rightarrow 0$. This property is
built in in the formula below, but in the main text, for simplicity we have ignored it. In this Appendix we show its consequences for the
measurables (cross sections and slopes).

We remind that the cross sections Eqs.~(\ref{SD}), (\ref{DD}) contain inelastic verteces with the kinematical factor:
 \begin{equation}\label{eq:inelVertex+Kf}
   {F_{inel}}^2(t,M_x^2)=A_{res}K_f(x_B,M_x^2)\sigma_T^{Pp}(M_i^2,t)+C_{bg}\sigma_{Bg},
 \end{equation}
 where $K_f(x_B,M_x^2)=\frac{x_B(1-x_B)^2}{(M_x^2-m^2) \left(1+\frac{4m^2x_B^2}{-t}\right)^{3/2}}$.

 Here $K_f(x_B,M_x^2)$ relates the nucleon structure function to the Pomeron-proton total crass section (for details see Ref.~\cite{PR}), $ \quad x_B=\frac{-t}{M_x^2-m_p^2-t}$ is the
Bjorken variable, that for high $M^2$ and moderate $t$ is $x\approx -t/M^2$, and we remind that $-t$ replaces the virtuality $Q$ in proton-Pomeron DIS, and $m_p$ is the proton mass. For large $s$ and small/moderate $t,\ \  x\approx -t/s.$ The Bjorken $x_B$ should not be confused with the Feynman variable $x_F$, defined by $1-x_F\approx (M_x^2-m^2)/x$.

Below we repeat some results if the main text, but with the above kinematical factor included.

With the kinematical factor included, some of the parameters should be refitted, namely:
$$A_{res}=15.5\,mb\cdot\mbox{~GeV}^2,\, b_{res}=-0.41\mbox{~GeV}^{-2},\, b_{bg}=-0.91\mbox{~GeV}^{-2}.$$ Other parameters are kept unchanged (see Tab. \ref{tab:ParSet}).

\begin{figure}[!ht]
 \includegraphics[width=0.49\linewidth,bb=13mm 24mm 195mm 185mm,clip]{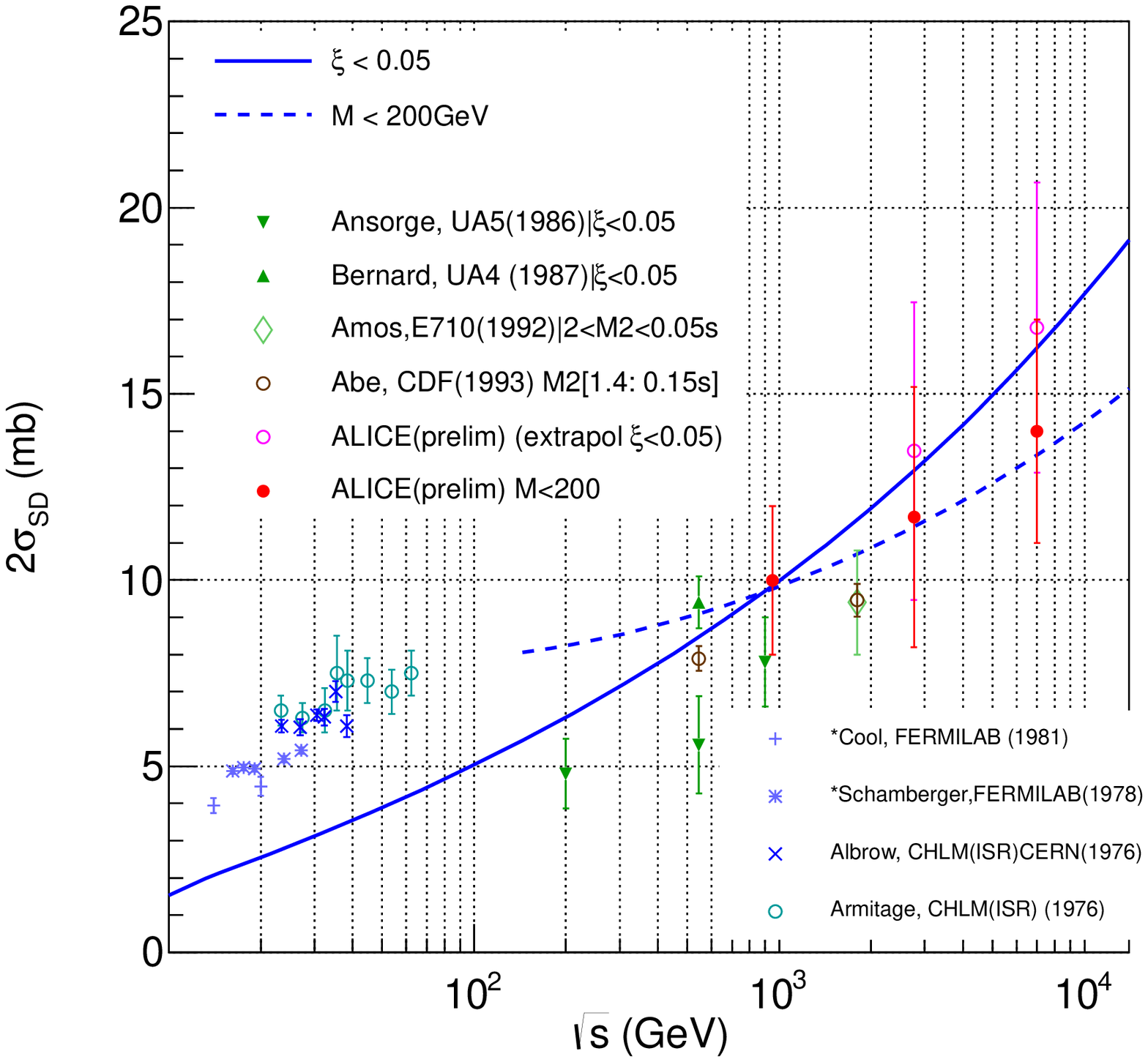}
 \includegraphics[width=0.49\linewidth,bb=13mm 24mm 195mm 185mm,clip]{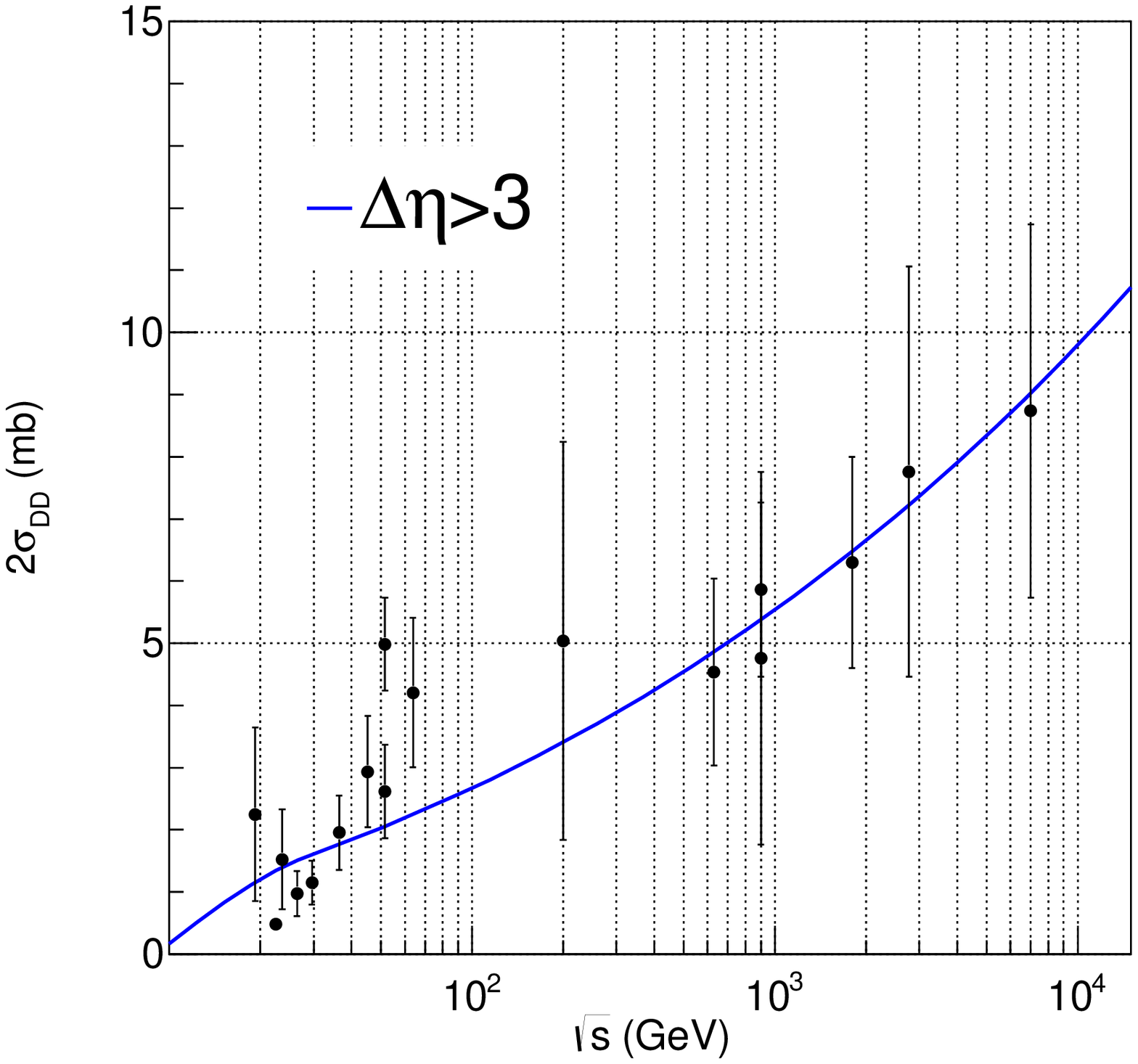}
 {(a)\hspace{0.47\linewidth}(b)}

 \caption{
 (a) SD cross section vs. energy $\sqrt{s}$ calculated from Eq.~(\ref{SD}) with the kinematical factor Eq.~(\ref{eq:inelVertex+Kf}). For $\sigma|_{\xi<0.05}$: $\chi^2/$n$=1.6$, n$=6$ (only $\sqrt{s}>100$~GeV); for $\sigma|_{M<200\mbox{~GeV}}$: $\chi^2/$n$=0.02$, n$=3$.\\
 (b) DD cross section vs. energy $\sqrt{s}$ calculated from Eq.~(\ref{DD}) with the kinematical factor Eq.~(\ref{eq:inelVertex+Kf}); $\chi^2/$n$=0.08$, $n=7$, for $\sqrt{s}>100$~GeV only.
  The experimental points are from \cite{Poghosyan for ALICE}. 
}
 \label{fig:cs.SD|cs.DD(+Kf)}
\end{figure}
\newpage

\begin{figure}[!ht]
 \includegraphics[width=0.49\linewidth,bb=13mm 24mm 195mm 185mm,clip]{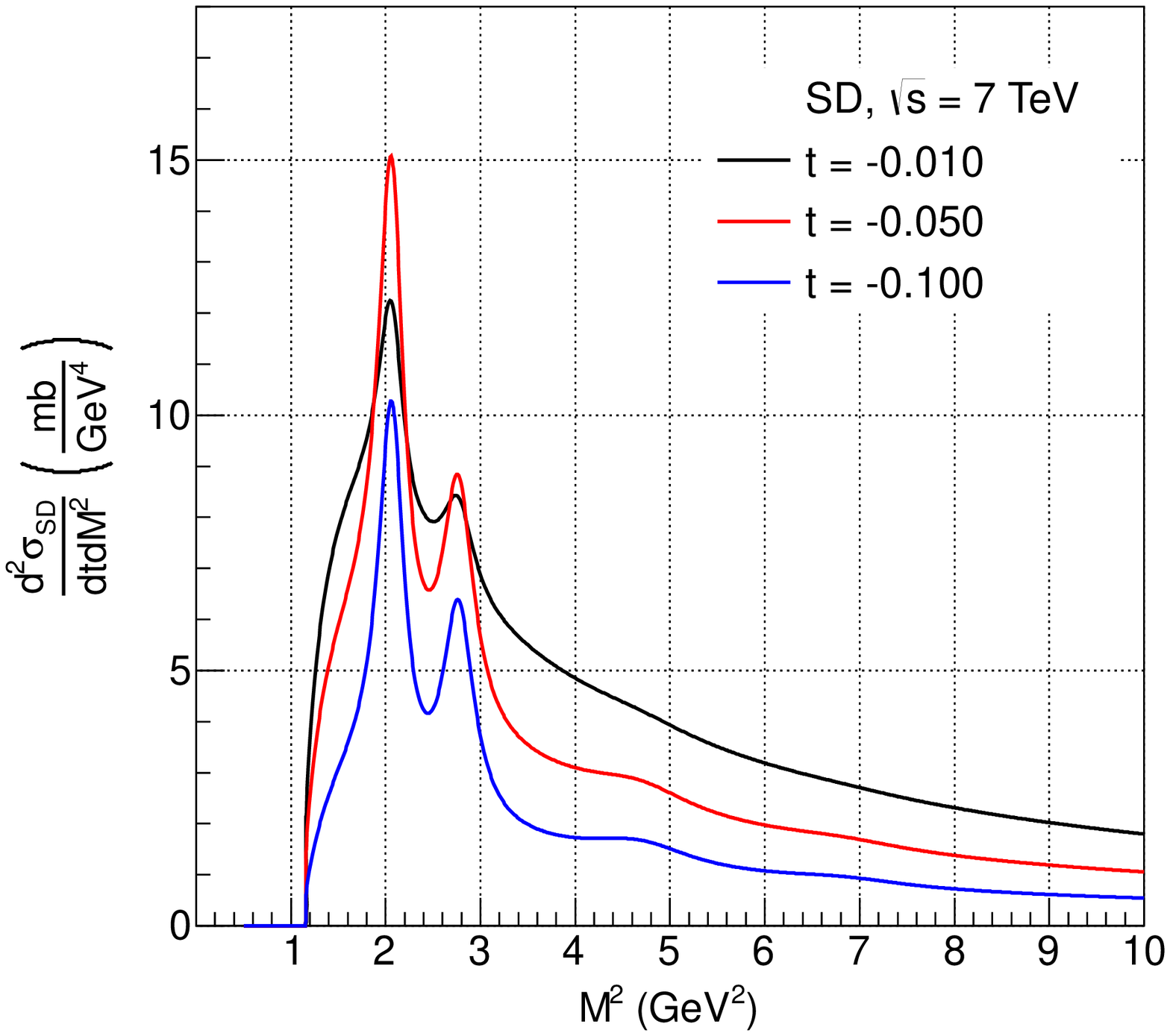}
 \includegraphics[width=0.49\linewidth,bb=13mm 24mm 195mm 185mm,clip]{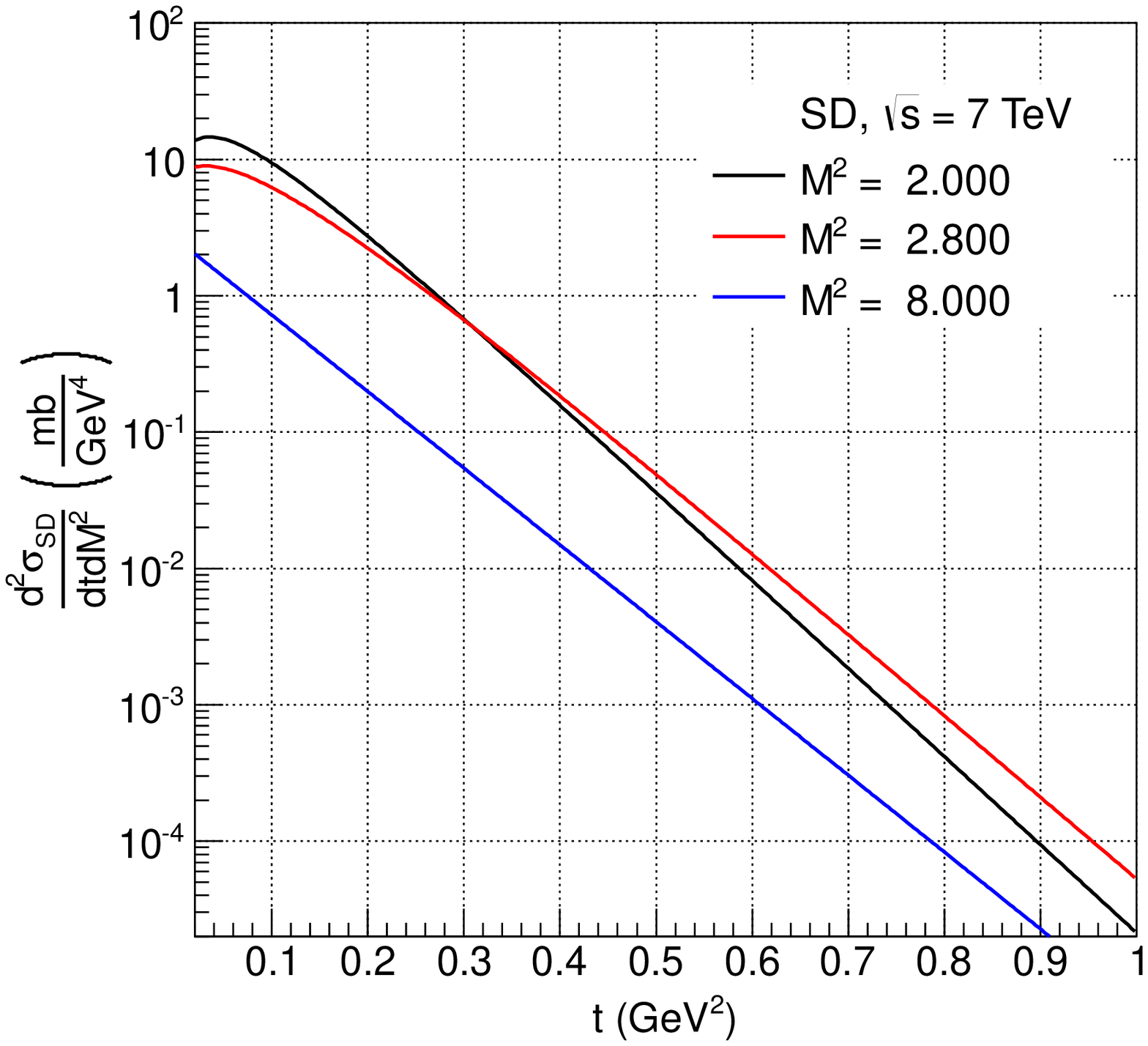}
  {(a)\hspace{0.47\linewidth}(b)}
 \caption{(a) Double differential SD cross sections as functions of $M^2$ for different $t$ values. \\
   (b) Double differential SD cross sections as functions of $t$ for different $M^2$ values.}
 \label{fig:d2cs.SD.Kf}
\end{figure}

\begin{figure}[!ht]
 \includegraphics[width=0.49\linewidth,bb=13mm 24mm 195mm 185mm,clip]{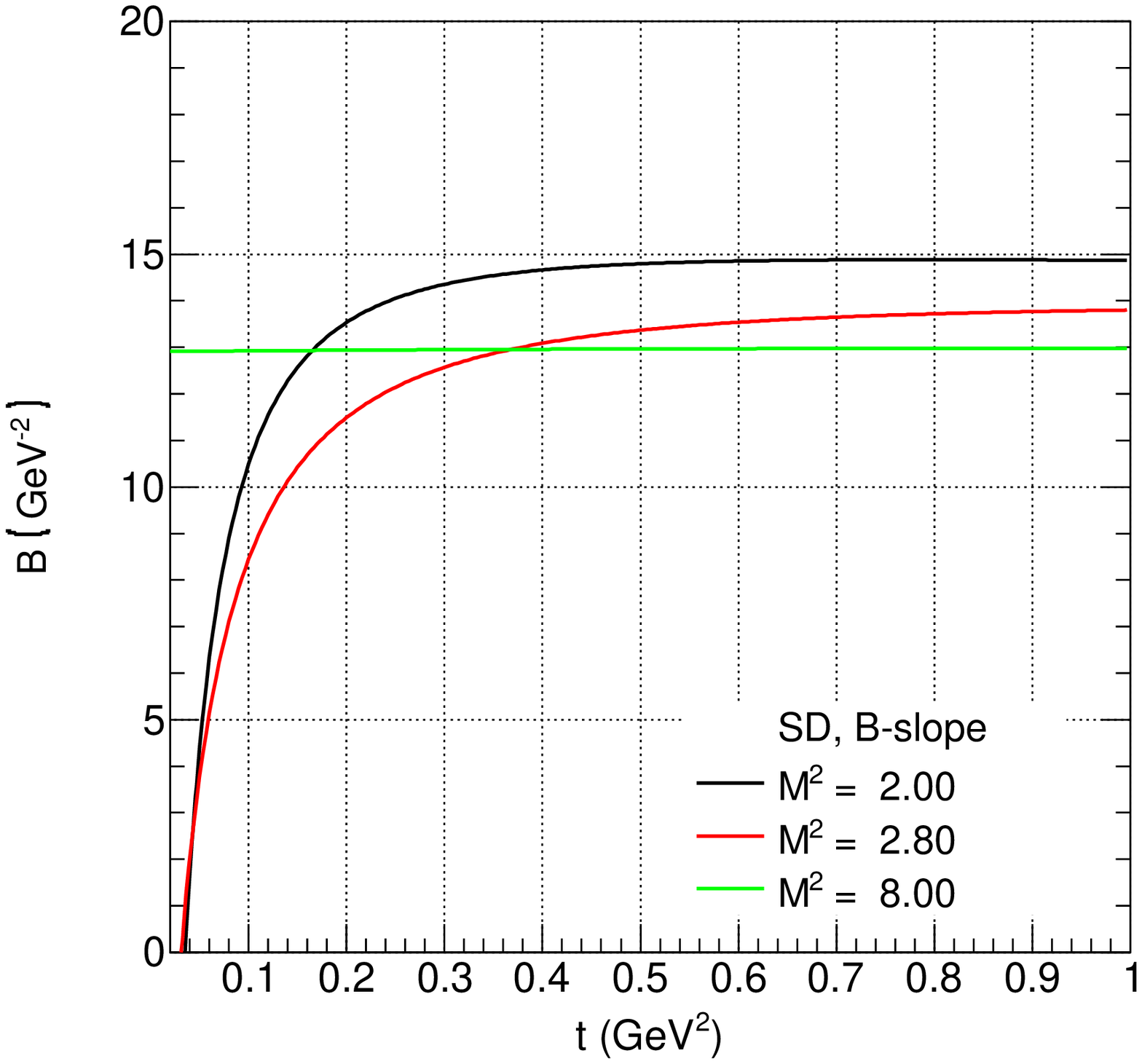}
 \includegraphics[width=0.49\linewidth,bb=13mm 24mm 195mm 185mm,clip]{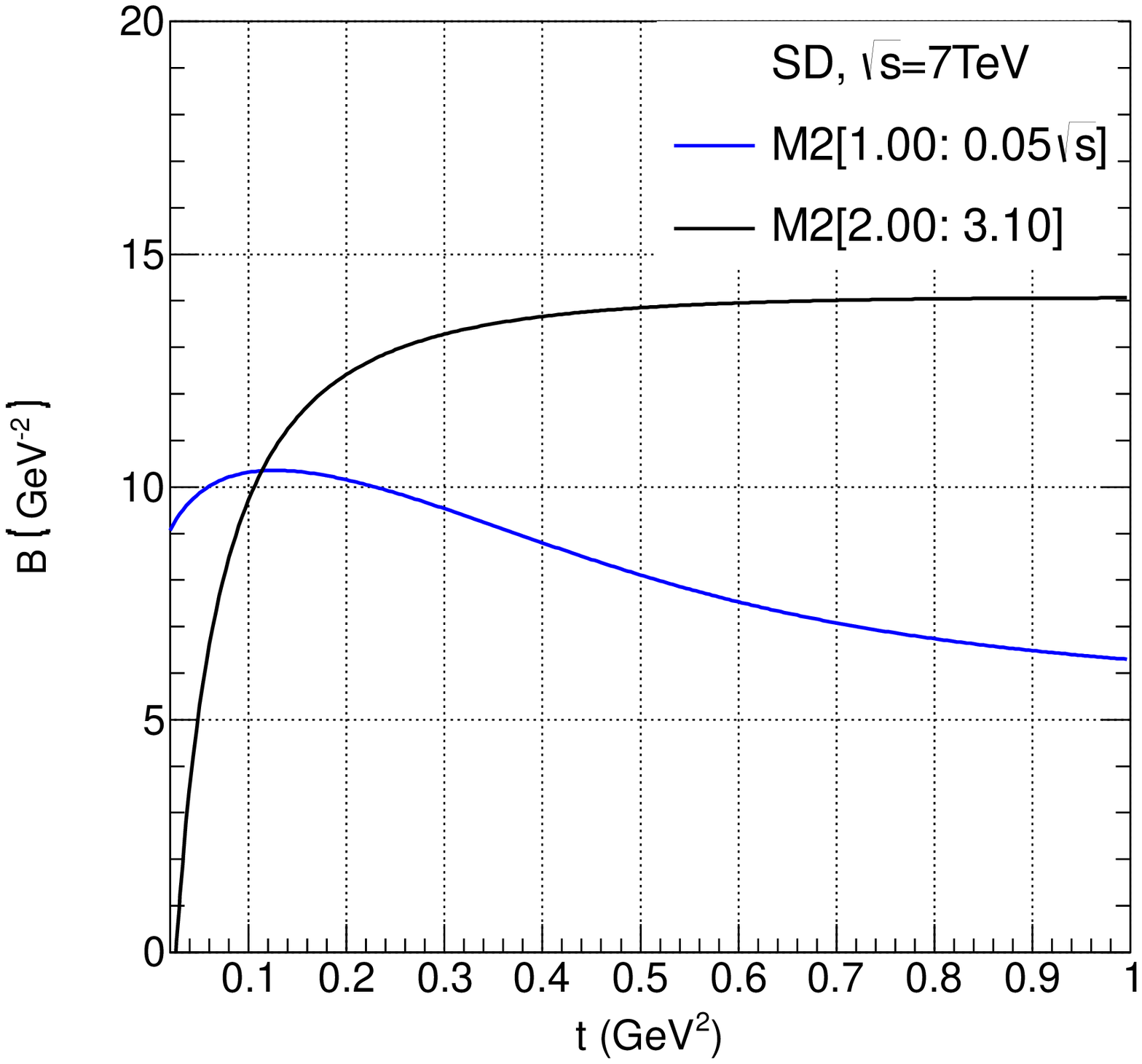}
  {(a)\hspace{0.47\linewidth}(b)}
 \caption{(a) Slopes of differential SD cross section at different $M^2$ values,
 $B=\frac{d}{dt}\ln\left.\frac{d^2\sigma_{SD}}{dtdM_x^2}\right|_{M^2_x=\{M^2_1, M^2_2, M^2_3\}}$.\\
 (b) Slopes of differential SD cross section integrated in $M_x^2$ over the region of the first resonance, and over the whole diffraction region. $B=\frac{d}{dt}\ln\frac{d\sigma_{SD}}{dt}$, where $\frac{d\sigma_{SD}}{dt}=\int_{M_1^2}^{M_2^2}\frac{d^2\sigma_{SD}}{dtdM_x^2}dM_x^2$.\\Double differential SD cross section $\frac{d^2\sigma_{SD}}{dtdM_x^2}$ is calculated from Eq.~(\ref{SD}) with the kinematical factor Eq.~(\ref{eq:inelVertex+Kf}) for $\sqrt{s}=7$~TeV.}
 \label{B.SD.Kf}
\end{figure}

\newpage
\begin{figure}[!ht]
 \includegraphics[width=0.49\linewidth,bb=13mm 24mm 195mm 185mm,clip]{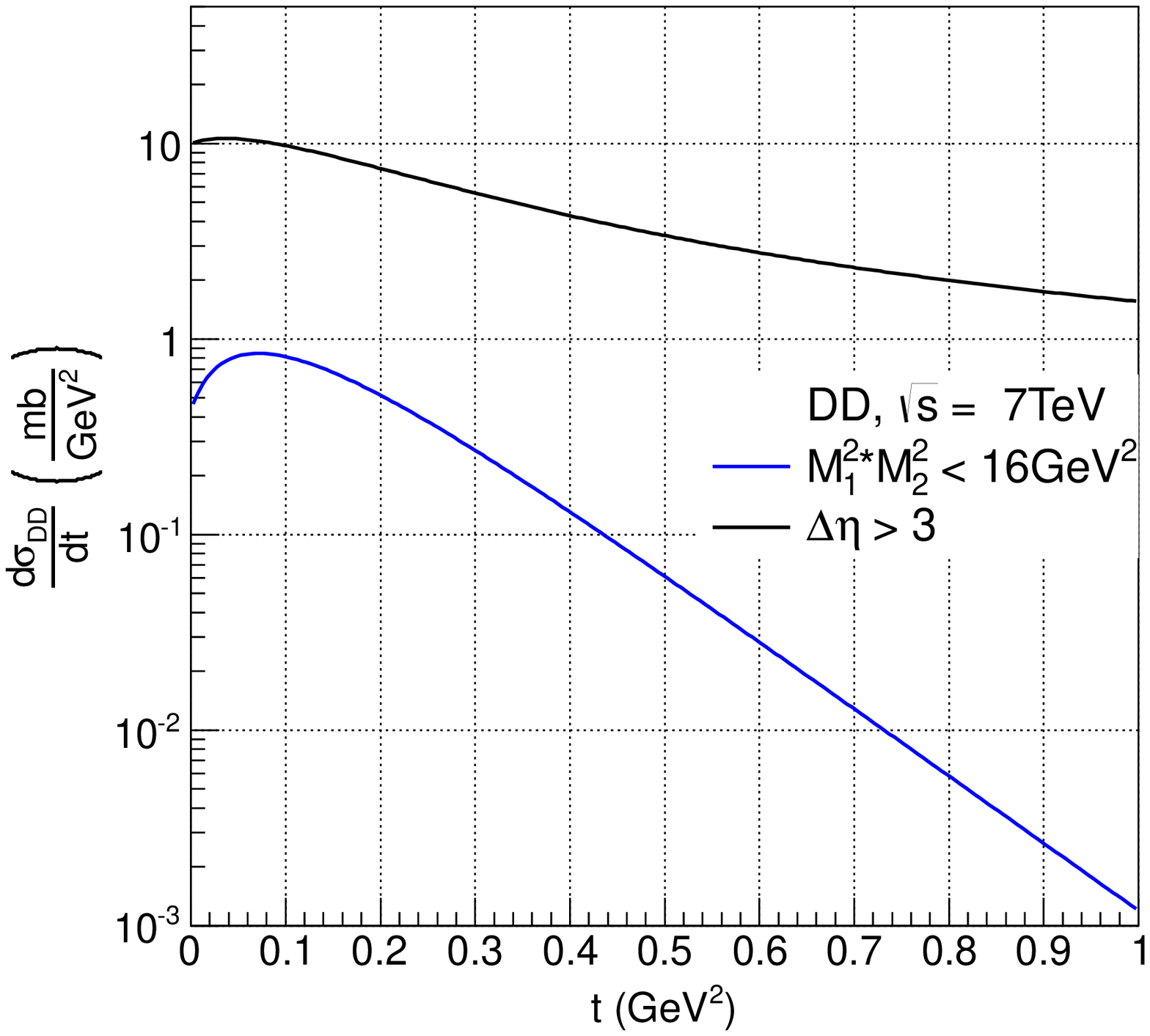}
 \includegraphics[width=0.49\linewidth,bb=13mm 24mm 195mm 185mm,clip]{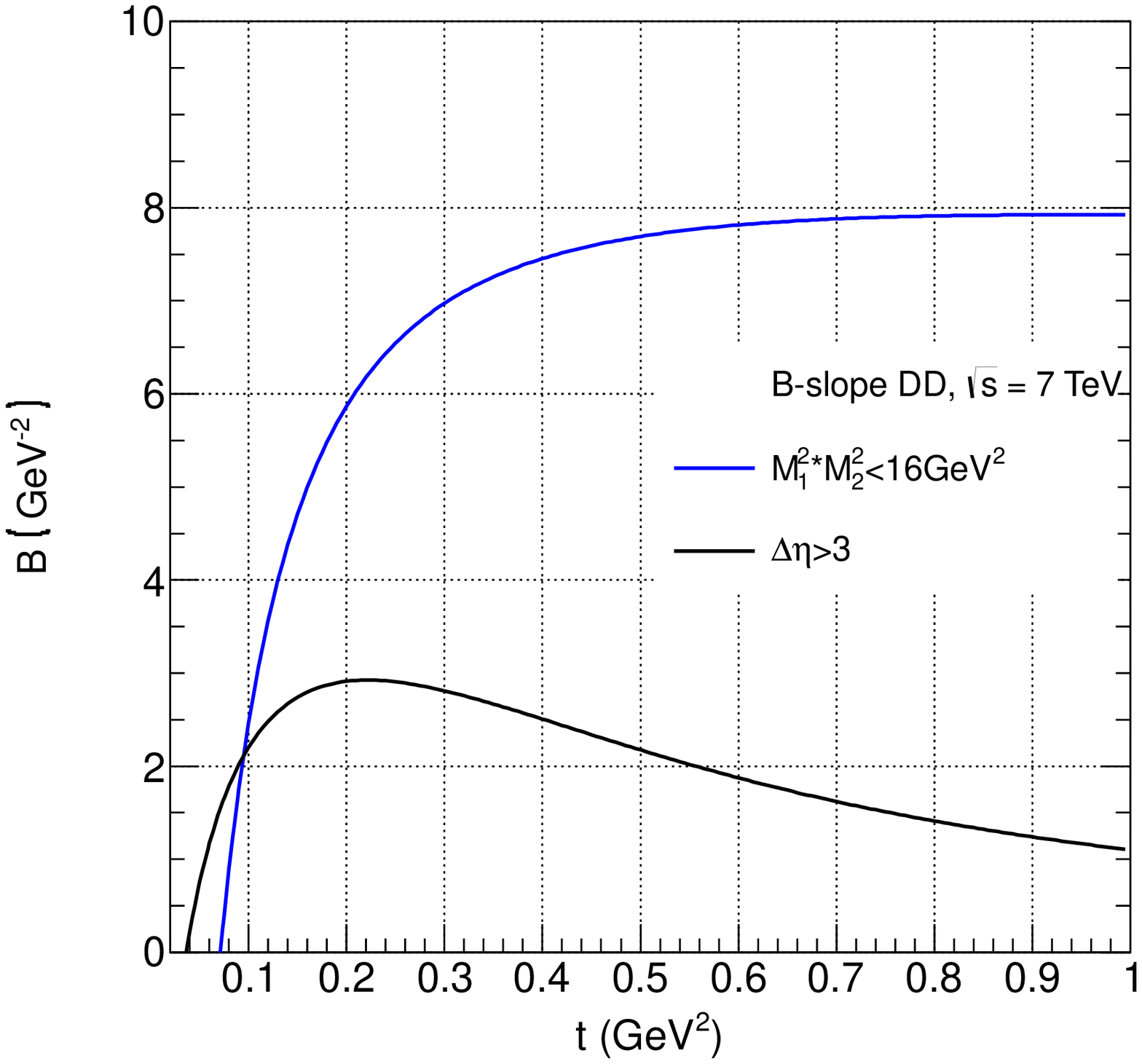}
  {(a)\hspace{0.47\linewidth}(b)}
 \caption{
 (a) Single differential cross section $\frac{d\sigma_{DD}}{dt}$ as a function of $t$ integrated over the region of resonances $\left(M_1^2M_2^2<16\mbox{~GeV}^2\right)$
 and over the whole region of diffraction ($\Delta\eta>3$).
 See Eq.~(\ref{DD})  with the kinematical factor Eq.~(\ref{eq:inelVertex+Kf});\\
 (b) The slope   $B=\frac{d}{dt}\ln\left(\frac{d\sigma_{DD}}{dt}\right)$ calculated from Eq.~(\ref{DD}) with the kinematical factor Eq.~(\ref{eq:inelVertex+Kf}).}
 \label{dcsdt|B.DD.Kf}
\end{figure}

\newpage
\section{Conclusions}\label{sec:Conclusions}

Below is a brief summary of our results:

\begin{itemize}
 \item At the LHC, in the diffraction cone region ($t<1$~GeV$^2$) proton-proton scattering is dominated (over $95 \%$) by Pomeron exchange (quantified in Ref.~\cite{JLL}). This enables full use of factorized Regge-pole models. Contributions from non-leading (secondary) trajectories can (and should be) included in the extension of the model to low energies, e.g. below those of the SPS.

 \item Unlike to the most of the approaches which use the triple Regge limit for construction of inclusive diffraction, our approaches based on the assumed similarity between the Pomeron-proton and virtual photon-proton scattering. The proton structure function (SF) probed by the Pomeron is the central object of our studies. This SF, similar to the DIS SF, is exhibits direct-channel (i.e. missing mass, $M$) resonances transformed in resonances in single- double- and central diffraction dissociation. The high-$M$ behaviour of the SF (or Pomeron-proton cross section) is Regge-behaved and contains two components: one decreasing roughly like $M^{-m},\ \ m\approx 2$ due to the exchange of a secondary Reggeon 
 (not to be confused with the Pomeron exchange in the $t$ channel!). The latter dominates the large-$M$ part of the cross sections. Its possible
manifestation may be seen in the data see Fig.~\ref{add.SD}(b). On the other hand, the large-$M$ region is the border
of diffraction, $\xi>0.05$.

 \item An important and intriguing prediction of the present model is the possible turn-down of the cross sections
towards $t=0$, see Appendix. The forward direction cannot be reached kinematically in SD or DD, moreover even the non-zero but small $|t|$ events are difficult to be reached, especially that they are masked by electromagnetic interactions (although weaker than in elastic scattering, see Ref.~\cite{GJS}. Further studies, both theoretical and experimental, of this intriguing phenomena are of great importance.

 \item The results of our calculations, that are mainly predictions for the LHC energies $7,\ 8$ and $14$~TeV, are collected in Table \ref{tab:cs.predict} Sec. \ref{sec:Results} .
%
%
%
%
Predicted values for integrated, within various limits of $t$ and/or $M^2$, cross sections can be also found in that section. The quality of the fit is quantified by the relevant $\chi^2$ values.

 \item The model has an important and interesting prediction, following from the gauge invariance of the structure functions ($Pp$ production amplitudes), namely that the cross sections turn down at very small values of $|t|$, probably accessible in the nearly forward direction of future measurements. This result was anticipated in 
    Ref.~\cite{PR}.

 \item Our approach in this paper is inclusive, ignoring e.g. the angular distribution of the produced particles from decaying resonances.
All resonances, except Roper, lie on the $N^*$ trajectory. Any complete study of the final states should included also spin degrees of freedom, ignored in the present model.

 \item For simplicity we used linear Regge trajectories and exponential residue functions, thus limiting the applicability of our model to low and intermediate values of $|t|$. Its extension to larger $|t|$ is straightforward and promising. It may reveal new phenomena, such as the the possible dip-bump structure is SD and DD as well as the transition to hard scattering at large momenta transfers, although it should be remembered that
diffraction (coherence) is limited (independently) both by $t$ and $\xi$.
\end{itemize}

\section*{Acknowledgements}
We thank K. Goulianos, V.K. Magas and R. Schicker for discussions. The work of L.J. was supported by the National Academy of Ukraine, under the grand ``Nuclear matter under extrim condition''.

\vfill \eject
\end{document}